\documentclass[twocolumn]{aastex63}
\usepackage{bm}
\usepackage{multirow}
\usepackage{amsmath}
\usepackage{color}

\newcommand{\hi}{\textsc{Hi}}

\newcommand{\mhalo}{\ensuremath{\textrm{M}_{\textrm{h}}}}
\newcommand{\mhi}{\ensuremath{\textrm{M}_{\textsc{Hi}}}}
\newcommand{\msun}{\ensuremath{\textrm{M}_{\odot}}}
\newcommand{\lgmhi}{\ensuremath{\log\mhi}}
\newcommand{\mstar}{\ensuremath{\textrm{M}_{\ast}}}
\newcommand{\lgmstar}{\ensuremath{\log\mstar}}

\shortauthors{Li et al.}

\begin{document}

\title{Conditional {\sc Hi} mass functions and the {\sc Hi}-to-halo mass relation in the local Universe}

\correspondingauthor{Xiao Li \& Cheng Li}

\author[0000-0002-2884-9781]{Xiao Li}
\affiliation{Department of Astronomy, Tsinghua University, Beijing 100084, China}
\email{li-x19@mails.tsinghua.edu.cn}

\author[0000-0002-8711-8970]{Cheng Li}
\affiliation{Department of Astronomy, Tsinghua University, Beijing 100084, China}
\email{cli2015@tsinghua.edu.cn}

\author[0000-0001-5356-2419]{H.J. Mo}
\affiliation{Department of Astronomy, University of Massachusetts Amherst, MA 01003, USA}

\author[0000-0003-1938-8669]{Ting Xiao}
\affiliation{Department of Physics, Zhejiang University, Hangzhou, Zhejiang 310027, China}

\author[0000-0002-6593-8820]{Jing Wang}
\affiliation{Kavli Institude for Astronomy and Astrophysics, Peking University, Beijing 100871,  China}

\begin{abstract}
	We present a new {H\sc{i}} mass estimator which relates
	the logarithm of the \hi-to-stellar mass ratio to a linear combination 
	of four galaxy properties: stellar surface mass density,
	color index $u-r$, stellar mass and concentration index, with 
	the scatter of individual galaxies around the mean \hi\ mass 
	modeled with a Gaussian distribution function. We calibrate the 
	estimator using the xGASS sample, including both \hi\ detection and
	non-detection, and constrain the model parameters through Bayesian 
	inferences.  Tests with mock catalogs demonstrate that our estimator 
	provides unbiased \hi\ masses for optical samples like the SDSS, 
	thus suitable for statistical studies of \hi\ gas contents 
	in galaxies and dark matter halos. We apply our estimator to the SDSS 
	spectroscopic sample to estimate the \hi\ mass function (HIMF) of local galaxies, 
	as well as the conditional \hi\ mass function (CHIMF) in galaxy groups
	and the \hi-halo mass (HIHM) relation. Our HIMF agrees with the ALFALFA 
	measurements at \mhi$\gtrsim 5\times10^9$\msun, but with higher amplitude and 
	a steeper slope at lower masses. We show that this discrepancy is 
	caused primarily by the cosmic variance which is corrected for the SDSS 
	sample but not for the ALFALFA. The CHIMFs for all halo masses can be 
	described by a single Schechter function, and this is true for red, blue and 
	satellite galaxies. For central galaxies the CHIMFs show a double-Gaussian 
	profile, with the two components contributed by the red and blue galaxies, 
	respectively. The total \hi\ mass in a group increases monotonically with halo mass.
	The \hi\ mass of central galaxies in galaxy groups increases rapidly 
	with halo mass only at \mhalo$\lesssim10^{12}$\msun, while the mass dependence 
	becomes much weaker at higher halo masses. The observed \hi-halo mass relation
	is not reproduced by current hydrodynamic simulations 
	and semi-analytic models of galaxy formation.
\end{abstract}


\section{Introduction} \label{sec:intro}

Galaxies are believed to form at the center of dark matter halos 
through gas cooling and condensation \citep{1978MNRAS.183..341W,MoBoschWhite2010}. 
Theoretical studies of the \hi-to-halo mass (HIHM) relation based on 
hydrodynamic simulations or halo-based models in recent years 
\citep{2013AJ....146..124H,2014MNRAS.440.2313B,	Guo2017,2018ApJ...866..135V,
	2018MNRAS.479.1627P,
	2019MNRAS.486.5124O,2020MNRAS.497..146D,2020MNRAS.498...44C,
	2021MNRAS.506.4893C,2021MNRAS.506.1507C,2022arXiv220710414L}
have suggested that the majority of the cold gas in the Universe, 
mostly in neutral (atomic) hydrogen (\hi) and molecular 
hydrogen (${\rm H}_2$), is expected to be in individual galaxies, 
with a small amount in the circum-galactic medium (CGM) and intergalactic 
medium (IGM). On the other hand, both theoretical studies 
and observations have shown that some of the cold gas may be thrown 
out of galaxies due to galactic winds driven by stellar and AGN 
feedback or environmental effects such as tidal stripping and 
ram-pressure stripping in massive halos. Clearly,  
measurements of \hi\ gas mass for large samples of galaxies 
down to small gas fractions and over large sky coverage
and redshift ranges are needed to provide stringent constraints 
on models of galaxy formation. 

Large surveys of \hi\ 21cm emission of galaxies have become 
available only in the past two decades, such as the \hi\ Parkes 
All-Sky Survey \citep[HIPASS;][]{2004MNRAS.350.1195M,2006MNRAS.371.1855W}
and the Arecibo Legacy Fast ALFA \citep[ALFALFA;][]{Giovanelli_2005}.
High spatial resolution surveys with interferometers are also becoming available, such as the APERture Tile In Focus array imaging survey \citep[Apertif;][]{Apertif_Adams2022}.
These surveys have detected \hi\ emission from tens of thousands 
of galaxies up to $z\sim0.06$, covering large areas in the sky. 
As expected, most (if not all) cases of the \hi\ detection are associated 
with optically-identified galaxies in the same sky position and redshift
\citep{2018ApJ...861...49H}. Both HIPASS and ALFALFA
have yielded measurements of the \hi\ mass function (HIMF) for the local
galaxy population over a mass range nearly five orders
of magnitudes from \mhi$\sim 10^6 \msun$ to \mhi$\sim 10^{11} \msun$
\citep{2005MNRAS.359L..30Z,2018MNRAS.477....2J}. The two measurements are quite similar to each other, though still with significant differences according to statistical errors, which might be caused by cosmic variance
\citep{Jones2018a,2019MNRAS.483.5334S}.
Compared to optical surveys, however, current \hi\ surveys are 
still shallow and biased to gas-rich galaxies. 
There have been recent efforts to measure the total  \hi\ mass 
in dark matter halos, e.g. by stacking ALFALFA data-cubes 
of galaxy groups of different halo mass \citep{2020ApJ...894...92G} 
or through deep observations of \hi\ emission of individual galaxies 
(Rhee et al. in preparation). \cite{2020ApJ...894...92G} found a positive 
correlation between the total {\hi} mass and the dark matter halo mass, 
with a significant excess of {\hi} mass at $M_h \sim 10^{12} \msun$, while
Rhee et al. (in preparation) observed a nearly flat relation. 
\cite{2021MNRAS.506.4893C} analyzed possible reasons for the discrepancy 
between these two studies, and showed that these measurements 
may not be able to obtain the intrinsic {\hi}-halo mass relation 
correctly, owing to uncertainties in the halo mass estimate and 
limitations of the stacking method.



  \hi\ samples with depths and sizes comparable to optical surveys, 
such as the Sloan Digital Sky Survey \citep[SDSS;][]{2000AJ....120.1579Y}, 
are difficult to obtain in general. As an alternative way to proceed, 
attempts have been made to estimate \hi\ masses for optically-selected 
galaxies using their spectral and/or photometric properties, 
based on the many observational evidences that the
galaxy \hi\ mass correlates with optical properties such as stellar mass 
\citep{Brown2015}, morphology 
\citep{Haynes1984,Toribio2011,2021MNRAS.505..304C} and optical size 
\citep{Haynes1984,Jones2018a}.
\citet{2004ApJ...613..898T} estimated surface gas mass densities 
for SDSS galaxies from surface densities of star formation rate 
(given by H$\alpha$ luminosities) by inverting the Kennicutt-Schmidt law 
\citep{1963ApJ...137..758S,1998ApJ...498..541K}. Meanwhile, 
\cite{2004ApJ...611L..89K} found that the \hi-to-stellar mass ratio 
($\log(\mhi/\mstar)$) of local galaxies is linearly correlated with
optical (e.g. $u-r$) and optical-to-near IR (e.g. $u-K$) colors, with a scatter of 
$\sim 0.4$ dex. Motivated by these studies, \cite{2009MNRAS.397.1243Z}
proposed an estimator for $\log(\mhi/\mstar)$ based on a linear 
combination of the $g-r$ color and $i$-band surface brightness 
$\mu_i$, with a scatter of $\sim0.3$ dex. They applied the estimator to 
the SDSS galaxy sample to investigate the variation of \hi\ gas 
mass fraction on the stellar mass-metallicity relation.
A follow-up study by the team of the GALEX Arecibo SDSS Survey
\citep[GASS;][]{2010MNRAS.403..683C} calibrated a similar estimator 
to link $\log(\mhi/\mstar)$ with $NUV-r$ and surface stellar mass 
density $\mu_\ast$, which also has a scatter of $\sim 0.3$ dex. 
\citet[][hereafter L12]{2012MNRAS.424.1471L} found that the two-parameter 
estimators from \cite{2009MNRAS.397.1243Z} and \cite{2010MNRAS.403..683C}
significantly underestimate the \hi\ mass for the ALFALFA sample, which 
is expected to be biased to \hi-rich galaxies. They proposed a new estimator 
with four galaxy properties: surface stellar mass density $\mu _\ast$, $NUV-r$, stellar mass 
$M_\ast$ and $g-i$ color gradient $\Delta_{g-i}$. With a smaller scatter of
$\sim0.2$ dex, this estimator was applied to the SDSS sample  
to study both the \hi\ mass-dependence of the galaxy clustering (L12) 
and environmental effects on the gas depletion in clusters of galaxies 
\citep{2013MNRAS.429.2191Z}. 

Using the REsolved Spectroscopy of a Local VolumE survey 
(RESOLVE; Kannappan et al. in prep.), \citet[]{2015ApJ...810..166E} calibrated 
a relation between the {\hi} gas fraction, color index and axis ratio of galaxies.
More recently, \citet[][hereafter Z20]{2020MNRAS.496..111Z} 
developed a two-parameter {H\sc{i}} mass estimator by linearly combining 
stellar mass $M_\ast$ and $(g-r)$, and used it to study the 
mass-metallicity relation of SDSS galaxies. To avoid the Malmquist bias 
produced by limited detection depths of current {H\sc{i}} surveys, Z20 designed a 
likelihood model to account for the detection probability of ALFALFA and 
constrained the model parameters using Bayesian inferences. 
\cite{2020arXiv200809804L} also used a linear combination of 
$M_\ast$ and $(g-r)$ to estimate the \hi\ mass of SDSS galaxies, 
in an attempt to estimate the {H\sc{i}} gas content in dark matter halos.
In addition to these linear estimators, a few nonlinear models have also 
been obtained to estimate \hi\ masses of local galaxies, taking advantage 
of the technique of machine learning 
\citep[e.g.][]{2017MNRAS.464.3796T,2018MNRAS.479.4509R}.

In this paper we extend previous studies by developing a new estimator
to predict the \hi\ gas content of galaxies in a large optical sample.
Our estimator makes improvements in the following aspects. First, we calibrate 
our \hi\ mass estimator using the xGASS sample \citep{xGASS}, and we 
include galaxies both with and without \hi\ detection in the calibration
to reduce bias. Recent studies have mostly used the ALFALFA sample, 
which is shallow and biased to \hi-rich galaxies. Galaxies without HI 
detection contain useful information about gas-poor galaxies, and 
so should not be ignored in the calibration in order to obtain an unbiased 
estimator. Second, we follow L12 to use a linear model using four galaxy 
properties for the estimator, but we also take into account scatter in the model. 
This scatter includes not only the intrinsic variance of individual galaxies
but also uncertainties in the measurements of the \hi\ mass and optical parameters.
Finally, we follow Z20 to constrain model parameters of our estimator 
using the Bayesian framework, which allows us to explore the model 
parameter space efficiently, and to better understand the correlations among 
model parameters. We use mock catalogs to mimic the selection effects of 
the ALFALFA and xGASS samples and to demonstrate that our estimator 
is unbiased for SDSS-like samples. We thus can use our estimator  
to estimate the HIMF of local galaxies, the conditional HIMF 
and the total \hi\ mass as a function of dark matter halo mass.

We organize our paper as follows. In \autoref{sec:data} we describe the
\hi\ and optical samples used in our analysis. In \autoref{sec:methods}
we present our {H\sc{i}} estimator and describe the calibration 
procedure and test results. In \autoref{sec:results}
we apply our {H\sc{i}} estimator to the SDSS sample to estimate {\hi} contents 
of galaxies and their host halos. We discuss and summarize our results 
in  \autoref{sec:discussion} and \autoref{sec:summary}.
Throughout this paper we assume a flat $\Lambda$CDM cosmology with 
$H_0=100h{\rm \ km\ s^{-1}Mpc^{-1}}$, $h=0.7$, $\Omega_m=0.3$,
and $\Omega_{\Lambda}=0.7$.

\section{Data} \label{sec:data}
\subsection{The xGASS sample} \label{sec:xGASS}

We use the xGASS sample to calibrate our \hi\ mass estimator. 
The xGASS is an extension of the GASS survey, an \hi\ 21cm survey  
observed with the Arecibo telescope for a sample of galaxies with 
redshift $0.025<z<0.05$ and a flat stellar mass distribution in 
the range $10^{10}\msun<\mstar<10^{11.5}\msun$. The sample galaxies 
are randomly selected from a parent sample of $\sim$ 12,000 galaxies 
located in the overlapping region among SDSS data release 
six \citep[DR6;][]{Adelman_McCarthy_2008}, 
Galaxy Evolution Explorer \citep[GALEX;][]{Martin_2005} Medium Imaging 
Survey, and the ALFALFA survey footprint. Each galaxy is observed with 
Arecibo until a significant \hi\ emission line is detected, or the \hi\ to
stellar mass ratio reaches an upper limit of $\mhi/\mstar\sim1.5\%$
(see C10 for details). The xGASS extends the survey down to a lower mass limit 
of $10^9 \textrm{M}_{\odot}$, by further observing a sample of galaxies with 
$9.0<\log ($\mhi$/$\msun$)<10.2$ and\footnote{For simplicity, we'll denote $\log_{10}$ as $\log$ throughout this paper} $0.01<z<0.02$. Here we use the
combined xGASS sample constructed by \cite{xGASS}, which includes 1179 galaxies   
obtained by GASS and xGASS and supplemented with {H\sc{i}}-rich galaxies 
selected from the ALFALFA $\alpha.70$ sample that are not included in GASS/xGASS. 

From the combined xGASS sample we select a subset of galaxies
with good measurements. A galaxy is included in our sample if 
all the following requirements are met: 
(i) $\tt{HIconf\_flag} \le 0$, (ii) $\tt{HI\_flag} \le 1$, and (iii) 
$\sigma _{u-r}<0.3$. Here $\tt{HIconf\_flag}$ is the confusion flag provided 
in the combined xGASS catalog, with $\tt{HIconf\_flag}=1$ for certain confusion, 
$>0$ for a small companion, $0$ for no confusion and $-99$ for non-detection. 
Thus, the first criterion selects both detections with no confusion 
and non-detections from the xGASS sample. The $\tt{HI\_flag}$ is the quality 
flag for xGASS detections, with $\tt{HI\_flag} \le 1$ for good quality, $2$ 
for marginal, $5$ for confused and $3$ for both marginal and confused observations. 
The error of $u-r$ color, $\sigma _{u-r}$, is calculated through error 
propagation based on the inverse variance of the $u-$ and $r-$band 
absolute magnitudes provided in the NASA Sloan Atlas \citep[NSA;][]{2011AJ....142...31B}.
These restrictions yield a final sample, which contains 625 galaxies 
each with an {H\sc{i}} mass measurement, and 358 galaxies each with only 
an {H\sc{i}} mass upper limit. For those with an \hi\ gas mass measurement, 
\mhi\ is self-absorption corrected using 
$\Delta \log M_{\rm {\hi}}=(0.13\pm 0.03)\log (a/b)$ \citep{2018MNRAS.477....2J},
where $a/b$ is the major-to-minor axis ratio. 
We refer to this sample as {\tt SampleX} in the rest of this paper.

\subsection{The ALFALFA galaxy sample} \label{sec:ALFALFA}

The Arecibo Legacy Fast Arecibo L-band Feed Array Survey 
\citep[ALFALFA;][]{Giovanelli_2005} is a blind survey of {H\sc{i}} 21 cm  
emission over $\sim7000\ {\rm deg^2}$ of the sky and up to 
redshift $z\sim0.06$. The final data release, the ALFALFA $\alpha.100$ 
catalog \citep{2018ApJ...861...49H}, consists of $\sim 31,500$ extragalactic 
{H\sc{i}} line sources. Two sky areas are covered by ALFALFA: one 
in the northern Galactic hemisphere and one in the southern Galactic hemisphere. 
Since our optical sample (see below) is mainly in the northern 
Galactic hemisphere, we select from ALFALFA $\alpha.100$ a subsample  
that covers the sky of $138^{\circ}<\alpha<232^{\circ},\ 0^{\circ}<\delta<36.5^{\circ}$ 
and the redshift of $0.00<z<0.05$. 
The upper limit in $z$ is set to avoid the contamination of RFI 
\citep{2018ApJ...861...49H}. 
We use ALFALFA sources flagged with ${\tt Code\ 1}$, i.e those with 
${\rm SNR \gtrsim 6.5}$, together with those flagged with ${\tt Code\ 2}$, 
i.e. those with ${\rm SNR\ (\lesssim 6.5)}$ but matched with optical 
counterparts that have redshift consistent with the observed 21cm line. 
Our ALFALFA galaxy sample consists of 16,400 galaxies. 
In what follows this sample is referred to as {\tt SampleA}.

\subsection{The observed HI mass function} \label{subsec:observed_HIMF}

\begin{figure}
    \centering
    \includegraphics[width=0.45\textwidth]{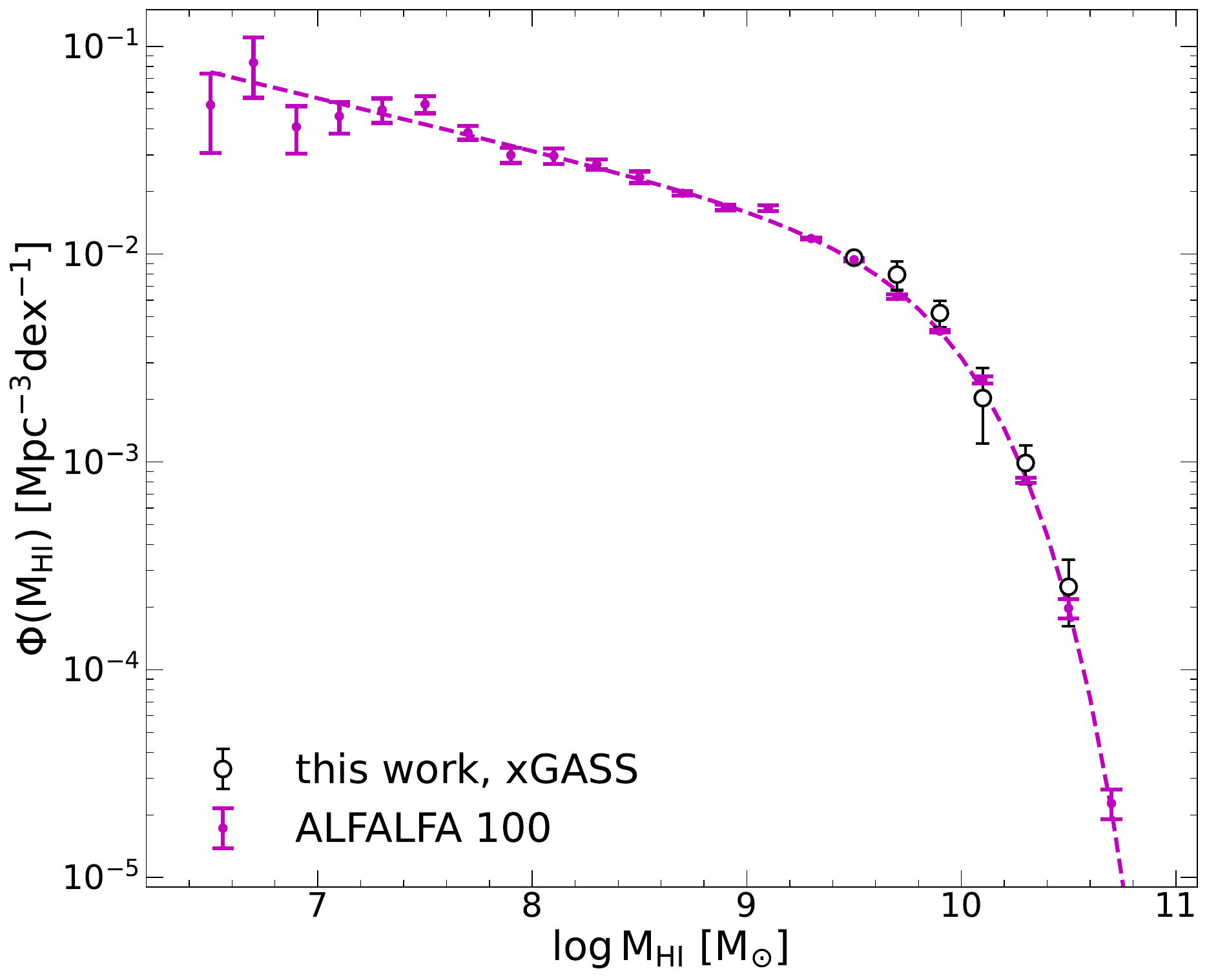}
    \caption{The HIMF of ALFALFA and xGASS survey. The magenta points with errorbar are the ALFALFA 100 percent HIMF datapoints \citep{2018MNRAS.477....2J}. The magenta dashed line
    is the best fit to the data. The black open circle with errorbar is the xGASS HIMF derived in this work (see \autoref{subsec:observed_HIMF}).}
    \label{fig:obs_HIMF}
\end{figure}

 A blind \hi\ survey, such as the ALFALFA, aims to detect all sources 
above some detection limit, e.g. in 21cm flux. Such a survey is expected 
to be biased against galaxies with low \hi\ mass. Thus, unlike an \hi\ survey
which starts from a given optical sample, such as xGASS described above, 
a blind survey does not provide a fair sample to study the relationship 
between \hi\ and optical properties of galaxies. However, being selected 
on the basis of 21cm flux, blind surveys can provide complete samples to 
obtain summary statistics, such as the \hi\ mass function (HIMF), above certain \hi\ mass.  
Such statistics are not direct measurements of the relationship between \hi\ and 
optical properties, they nevertheless provide constraints on the relationship. 
In our analysis, we will use the HIMF obtained from ALFALFA as an additional 
constraint on our model. 

As a check of consistency between xGASS and ALFALFA, we show in \autoref{fig:obs_HIMF}
the HIMFs obtained from these two samples. The ALFALFA measurements
are adopted from \cite{2018MNRAS.477....2J} and shown as the magenta error bars, 
together with the fitting to a Schechter function shown by the magenta dashed line. 
The HIMF for xGASS is estimated by us using the xGASS representative sample 
({\tt SampleX}). To that end, we assign a weight $w_{i1}w_{i2}$ to the $i$-th galaxy 
in {\tt SampleX}, where $w_{i1}$ accounts for the mass-dependent selection 
of {\tt SampleX} with respect to the SDSS galaxy sample 
({\tt SampleS}), and $w_{i2}$ corrects for the  selection effect of 
{\tt SampleS} using the $1/V_{\rm max}$ weighting scheme
(see \autoref{sec:HIMF} for more details).
We only obtain the xGASS HIMF at $\log M_{\rm {\hi}}/M_{\odot}>9.5$ to avoid 
influences of non-detection and incompleteness at the low-mass end. 
The errors of the HIMF are estimated by bootstrap resampling of the 
galaxies in {\tt SampleX}. As one can see from the figure, 
at $\log M_{\rm {\hi}}/M_{\odot}>9.5$, both the xGASS and ALFALFA measurements 
are consistent with each other. To ensure consistency, we only use the 
ALFALFA measurements of the HIMF at $\log M_{\rm {\hi}}/M_{\odot}>9.5$ 
as additional constraints on our \hi\ estimator.

\subsection{The SDSS galaxy sample} \label{sec:SDSS}

We select our optical galaxy sample from the New York University 
Value Added Galaxy Catalog (NYU-VAGC)\footnote{http://sdss.physics.nyu.edu/vagc/} 
constructed by \cite{Blanton_2005} from the SDSS spectroscopic galaxy sample. 
We start with the VAGC post-redshift sample {\tt bbright0}, which contains 
535,192 galaxies with spectroscopically measured redshifts and brighter than 
$r=17.6$ mag, where $r$ is the $r$-band apparent Petrosian magnitude corrected 
for Galactic extinction. We restrict ourselves to the redshift range of 
$0.003<z<0.05$, where the upper redshift limit is set to match the \hi\ galaxy 
sample and the lower limit is set to avoid large uncertainties in 
distances produced by peculiar motions of nearby galaxies. 
This gives us a sample of 86,487 galaxies.
We refer to this sample as {\tt SampleS} in the rest of this paper.
We will apply our \hi\ mass estimator to this sample and  
derive the statistical properties of {H\sc{i}} gas contents for both 
galaxies and their host dark matter halos (see \autoref{sec:results}).

\subsection{The NSA and galaxy properties}
\label{sec:properties}

To design our \hi\ mass estimator, we obtain, for each galaxy in the xGASS, 
ALFALFA and SDSS samples described above, the following set of properties
from the NSA\footnote{http://www.nsatlas.org/}:
\begin{itemize}
\item \mstar: stellar mass in units of solar mass, estimated 
with ${\tt kcorrect\ v4\_2}$,\footnote{http://kcorrect.org} \citep{2007AJ....133..734B}
which performs SED fitting to the SDSS photometric data based on the 
        stellar spectral templates from \cite{2003MNRAS.344.1000B} using the stellar 
        initial mass function of \cite{2003PASP..115..763C}
        and the Padova 1994 isochrones, and the ionized gas emission spectral 
        templates of \cite{2001ApJ...556..121K}.
\item $\mu_\ast$: the surface stellar mass density, defined as $M_*/(2 \pi R_{50}^2)$,
        where $R_{50}$, in units of kpc, is the radius enclosing a half of the total light in $r$-band.
\item $u-r$: the $u-r$ color index, given by the K-corrected absolute magnitude 
difference between $u$- and $r$-band using elliptical Petrosian photometry of SDSS images.
\item $NUV-r$: the $NUV-r$ color index, where $NUV$ is the K-corrected 
magnitude in the $NUV$-band using elliptical Petrosian photometry of GALEX images.
\item $R_{90}/R_{50}$: the concentration index defined as the ratio between 
$R_{90}$ (the radius enclosing 90\% of the total light in $r$-band) and $R_{50}$.
\end{itemize}

\begin{figure}
    \includegraphics[width=0.5\textwidth]{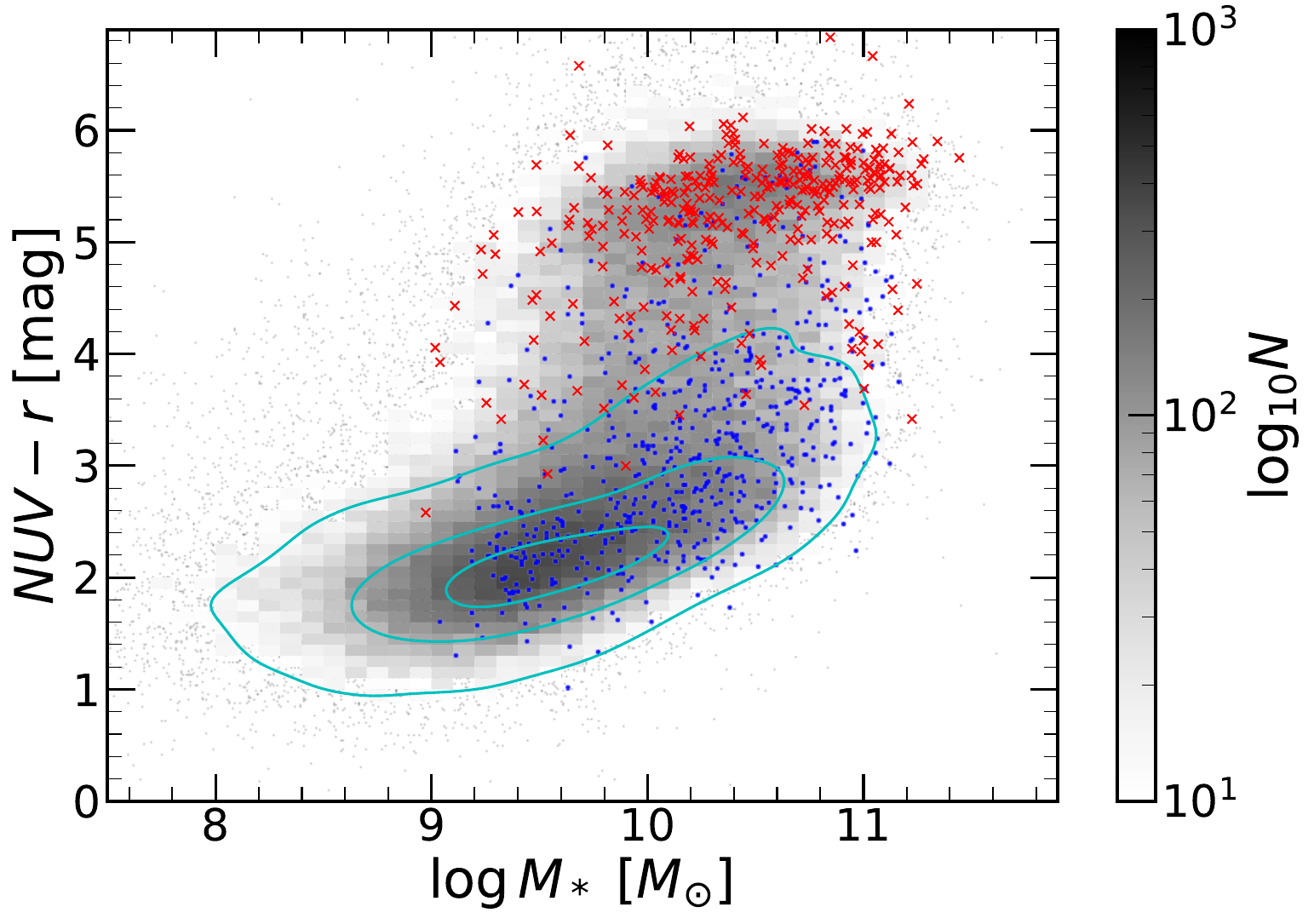}
    \caption{The distribution of our galaxy samples in NUV-r color and stellar mass diagram.
            Blue points are {H\sc{i}}-detected galaxies in \texttt{SampleX} and
            red crosses indicate \hi\ non-detections in \texttt{SampleX}.     
            Cyan contours indicate the distribution of \texttt{SampleA} galaxies. 
            From inside to outside, the contours include 16\%, 50\% and 84\% of the \texttt{SampleA} respectively. The grey 2D histogram and dots show the distribution of \texttt{SampleS} galaxies.}

    \label{fig:data_points}
\end{figure}

About 5\% of the galaxies in {\tt SampleS} do not have counterparts 
in the NSA. We have examined the effects of these missing galaxies on our results 
and found that they are negligible. Figure \ref{fig:data_points} 
displays the distribution of the \texttt{SampleX}, \texttt{SampleA} and \texttt{SampleS} 
in the plane of $NUV-r$ color and stellar mass. Galaxies with \hi\ detection and 
non-detection in \texttt{SampleX} are plotted as blue dots and red crosses respectively,
respectively. The cyan contours represent the distribution of 
\texttt{SampleA}. The grey 2D histogram and dots indicate \texttt{SampleS}.
We can see that \texttt{SampleA} galaxies are mainly located in the 
star-forming sequence with $NUV-r\lesssim4$. Compared to \texttt{SampleA}, 
\texttt{SampleX} covers a larger region in parameter space (although it is 
limited to \mstar$>10^9$\msun\ and with a smaller sample size), thus can better 
represent the general galaxy population. Therefore, we use \texttt{SampleX} to 
calibrate our \hi\ mass estimator.

\section{Predicting the {H\sc{i}} mass of galaxies} \label{sec:methods}

\subsection{The \hi\ mass estimator}
\label{sec:hi_estimator}

Consider a sample of $N_g$ galaxies indexed by $i$, each of which has  
a set of observational parameters, $\bm{x}_i$, in addition to the \hi\ 
observation, $y_i$. We aim to calibrate a general relation between $y$
and $\bm{x}$ so that it can be used to reliably predict $y$ from $\bm{x}$ 
for galaxies in other samples. We model this relation as
\begin{equation}\label{eqn:model}
  y (\bm{x}) = y_0(\bm{x}) + \delta y(\sigma),
\end{equation}
where $y_0$ gives the mean value of $y$ for galaxies of given $\bm{x}$, 
and $\delta y$ quantifies the scatter of individual galaxies around the mean. 
$\delta y$ is assumed to be a random variable that follows a Gaussian distribution 
function with a zero mean and a dispersion of $\sigma$. We model the mean
of $y$ as a linear combination of $\bm{x}$:
\begin{equation}\label{eqn:y0}
  y_0(\bm{x}) = \bm{\beta}^{T}\bm{x}+q,
\end{equation}
where $\bm{\beta}$ is a vector of model parameters that has the same number 
of elements as $\bm{x}$, $T$ denotes transpose, and $q$ is a constant. In this work we choose $y$ to be the \hi-to-stellar mass 
ratio in logarithmic scale, $y= \log(\mhi/\mstar)$, and
We use {\tt SampleX} to calibrate model parameters. 
For $\bm{x}$, we adopt the following four parameters: the 
surface stellar mass density $\mu_\ast$, the color index $u-r$, the 
stellar mass \mstar\ and the concentration index 
$R_{90}/R_{50}$ (see \S~\ref{sec:properties}). Thus
\begin{equation}\label{eqn:yxbeta}
y = \log\frac{\mhi}{\mstar},
~~~
\bm{x} = \left(
        \begin{aligned}
            & \log \mu_\ast \\
            & u-r \\
            & \log \mstar\\
            & \log\frac{R_{90}}{R_{50}}
         \end{aligned}
    \right),
    ~~~
    \bm{\beta} = \left(
    \begin{aligned}
       & a \\
       & b \\
       & c \\
       & d  
    \end{aligned}    
    \right).
\end{equation}
It is not straightforward  to design an appropriate functional form for $\sigma$. 
Our tests showed that the scatter actually 
depends on $m_0\equiv y_0+\lgmstar$, that is, the logarithm of 
the mean value of \hi-to-stellar mass ratio, as given by the \hi\ estimator, plus the logarithm of the stellar mass.
The dependence can be described by a piecewise linear function: 
\begin{equation}\label{eqn:sigma}
	\sigma(m_0) = \left\{
	\begin{aligned}
		& |c_a m_0 + c_b|, & \textrm{if}~ m_0 \geq m_{0,t} \\
		& |c_a m_{0,t} + c_b|, & \textrm{if}~ m_0 < m_{0,t} 
	\end{aligned}
	\right.
\end{equation}
where $c_a$ and $c_b$ are model parameters to be determined, and 
$m_{0,t}$ is an empirically determined cut in $m_0$ below which the 
scatter is a constant.
Our model thus has a total of 7 free parameters: 
$a,\ b,\ c,\ d,\ q,\ c_a,\ c_b$, to be determined.
The four-parameter model for the mean relation (\autoref{eqn:y0})  
is motivated by that in L12
where the \hi-to-stellar mass ratio of galaxies was considered as a function of 
$\mu_\ast$, $NUV-r$, \mstar\ and the radial gradient of $g-i$. 
Here we replace the $g-i$ color gradient by the concentration index
which is more widely used and easier to measure. We also replace 
$NUV-r$ by $u-r$ for similar reasons. We note that our model and
that of L12 lead to similar results in terms of estimating 
the mean relation between \hi\ gas mass and galaxy properties. 
We have attempted to include more parameters or to replace some of 
the parameters, and found no significant improvements. As pointed out in 
the introduction, our \hi\ mass estimator is an improvement 
on earlier estimators because of the inclusion of both the scatter in the model 
and non-detections in the calibration sample.

\subsection{Bayesian inferences of model parameters}
\label{sec:bayes_likelihood}

We make Bayesian inferences for the model parameters of our \hi\ mass 
estimator. In what follows we denote the set of model parameters 
as $\bm{\theta}\equiv\{a,b,c,d,q,c_a,c_b\}$. 
We use a combination of two sets of observational data.
The first set includes measurements of the four galaxy properties 
and the \hi-to-stellar mass ratio for all individual galaxies in {\tt SampleX},
and is denoted as $\bm{D}\equiv\{\bm{x}_i,y_i\}$ ($i=1,...,N_g$), where $N_g$ 
is the number of galaxies in the sample. 
The second set is the HIMF estimated from ALFALFA, and is denoted as 
$\Phi_\hi\equiv\Phi(\mhi)=\{\Phi_{\hi,j}\}$ ($j=1,...,N_m$), with $N_m$ the 
number of mass bins of the HIMF. Applying Bayes Theorem, we can write 
the likelihood of $\bm{\theta}$ given both $\bm{D}$ and $\Phi_\hi$ as 
\begin{equation}\label{eqn:bayes_both}
	\begin{aligned}
		P(\bm{\theta}|\Phi_\hi,\bm{D})  
		& =  
		\frac{P(\Phi_\hi,\bm{D}|\bm{\theta})\cdot P(\bm{\theta})}{P(\Phi_\hi,\bm{D})}	 \\ 
		& =  
		\frac{P(\Phi_\hi|\bm{\theta})\cdot P(\bm{D}|\bm{\theta})\cdot P(\bm{\theta})}
		{P(\Phi_\hi)\cdot P(\bm{D})} \\ 
		& = 
		\frac{P(\Phi_\hi|\bm{\theta})\cdot P(\bm{\theta}|\bm{D})}{P(\Phi_\hi)},
	\end{aligned}
\end{equation}
where in the second line we have assumed that the measurements of $D$ and $\Phi_\hi$ are independent. 
The third line follows from the Bayes relation, 
\begin{equation}\label{eqn:bayes_d}
	P(\bm{\theta}|\bm{D})=\frac{P(\bm{D}|\bm{\theta})\cdot P(\bm{\theta})}{P(\bm{D})}
\end{equation}
where $P(\bm{\theta}|\bm{D})$ is the posterior distribution of $\bm{\theta}$
inferred from $\bm{D}$ alone. 
The above equations show that we can derive $P(\bm{\theta}|\Phi_\hi,\bm{D})$ 
in two steps. First, we obtain $P(\bm{\theta}|\bm{D})$ from  
$\bm{D}$ using an assumed prior distribution $P(\bm{\theta})$. 
Second, we infer $P(\bm{\theta}|\Phi_\hi,\bm{D})$ from $\Phi_\hi$ using 
$P(\bm{\theta}|\bm{D})$ as the prior. 

\begin{figure}[t!]
    \includegraphics[width=0.47\textwidth]{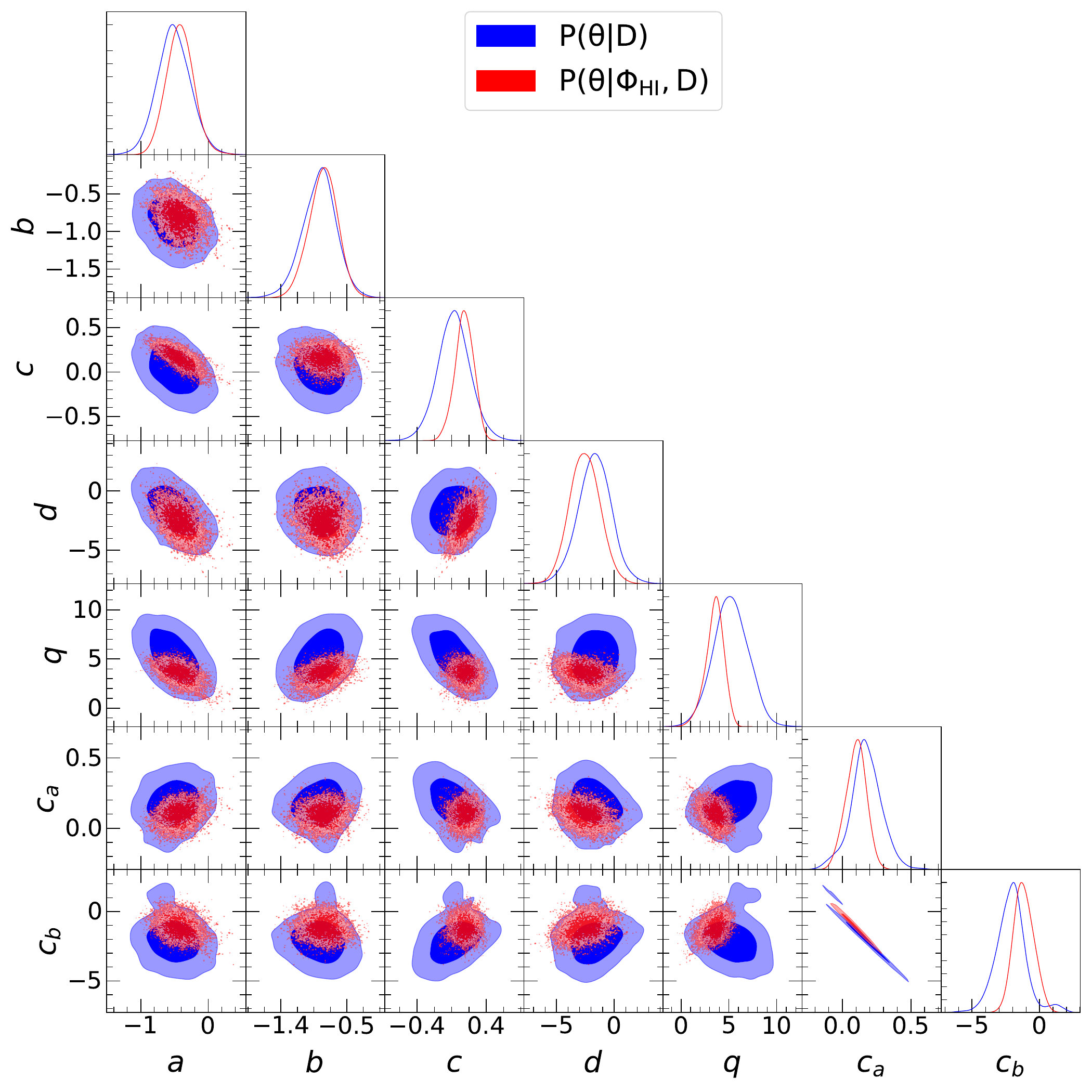}
    \caption{Probability distribution of model parameters. Contours correspond to 1-$\sigma$ and 2-$\sigma$ from inside out.
    The blue contours show the distribution of $P(\bm{\theta} | \bm{D})$, 
	and the red contours show the distribution of $P(\bm{\theta} | \Phi_{\hi},\bm{D})$.}
    \label{fig:posterior}
\end{figure}

We start by deriving $P(\bm{\theta}|\bm{D})$ using \autoref{eqn:bayes_d}.
Assuming that galaxies are sampled with an inhomogeneous Poisson process,
we can write \autoref{eqn:bayes_d} in logarithmic form as
\begin{equation}\label{eqn:likelihood}
	\begin{aligned}
	\ln P(\bm{\theta}|\bm{D})= & \ln P(\bm{D}|\bm{\theta}) + \ln P(\theta) 
	- \ln P(\bm{D})  \\ 
	                         = & \sum_{i\in \textrm{S}_{\textrm{d}}}
	                          \ln P_{\textrm{d},i}(y_i|\bm{\theta},\bm{x}_i)
	                          + \sum_{i\in \textrm{S}_{\textrm{n}}}
	                          \ln P_{\textrm{n},i}(y_i|\bm{\theta},\bm{x}_i) \\ 
	                          & + \textrm{const.}
	\end{aligned}
\end{equation}
Here we have assumed a flat prior distribution for all model parameters, so that 
$\ln P(\bm{\theta})$ and the data distribution $P(\bm{D})$ both become constant
and can be ignored. In the second equation $\textrm{S}_\textrm{d}$ and 
$\textrm{S}_\textrm{n}$ denote the subsets of galaxies in {\tt SampleX} 
with and without \hi\ detection, respectively,
while $P_{\textrm{d},i}$ and $P_{\textrm{n},i}$ are the corresponding 
likelihoods of $y_i$ given $\bm{\theta}$ and $\bm{x}_i$.
The likelihood for the $i$-th galaxy to have an \hi\ detection is modeled by 
\begin{equation} \label{eqn:det_likelihood}
    P_{\textrm{d},i}(y_i|\bm{\theta},\bm{x}_i)=
    \frac{1}{\sqrt{2\pi}\sigma_i}\exp\left\{-\frac{(y_{i}-y_{0,i})^2}{2\sigma_i^2}\right\},
\end{equation}
where $y_{0,i}=y_{0,i}(\bm{x}_i)$ is the mean value predicted by \autoref{eqn:y0},
and $\sigma_i=\sigma_i(m_{0,i})$ is the scatter given by \autoref{eqn:sigma}.
When calculating $\sigma_i$, we set the value of $m_{0,t}$ in \autoref{eqn:sigma}  
to be $m_{0,t}=8.5$ based on tests which show that our results are 
insensitive to the exact value of $m_{0,t}$. 
We note that, for simplicity, we have neglected the measurement 
error of $\bm{x}$ in \autoref{eqn:det_likelihood}, which we find to cause a 
difference of only $\sim 0.01$ dex in the predicted \hi\ mass of our galaxies 
and thus have little effect on our results.
For non-detections with only upper limits of $y_i$, we assume 
they follow the same conditional distribution as detections, so that the 
likelihood $P_{\textrm{n},i}$ can be calculated by integrating  $P_{\textrm{d},i}$ 
over $y\leq y_i$, i.e.
\begin{equation}
    P_{\textrm{n},i}(y_i|\bm{\theta},\bm{x}_i) = 
    \int^{y_i}_{-\infty} P_{\textrm{d},i}(y^\prime|\bm{x}_i,\bm{\theta})dy^\prime\,.
\end{equation}
We use the Python package $\tt{emcee}$ to perform Markov chain Monte Carlo (MCMC)
sampling over the parameter space \citep{emcee}, and obtain the posterior 
distribution of the model parameters using \autoref{eqn:likelihood}. During 
this process we weight each galaxy in {\tt SampleX} in such a way that the 
stellar mass distribution estimated by the weighted galaxies matches the stellar 
mass function (GSMF) in the local universe as measured by
 \cite{2019ApJ...872..180C} (see their Table 1). 
The posterior distributions of all the model parameters obtained this way 
are plotted in \autoref{fig:posterior} as blue contours. 

Next, we use the ALFALFA HIMF $\Phi(\mhi)$ at $\log M_{\rm {\hi}}/M_{\odot}>9.5$ 
to further constrain our model parameters, 
adopting the likelihood derived above, $P(\bm{\theta}|\bm{D})$, as the prior. 
Again, we write the last equation in \autoref{eqn:bayes_both} in 
logarithmic form as
\begin{equation}\label{eqn:model2}
	\begin{aligned}
		\ln P(\bm{\theta}|\Phi_\hi,\bm{D})  
		= & \ln P(\Phi_\hi|\bm{\theta}) +  \ln P(\bm{\theta}|\bm{D}) 
		+ \textrm{const.}  \\
		= & \sum_j \ln P_j(\Phi_{\hi,j}|\bm{\theta}) +  \ln P(\bm{\theta}|\bm{D}) 
		+ \textrm{const.},
	\end{aligned}
\end{equation}
where $\Phi_{\hi,j}$ is the HIMF measurement in the $j$-th mass bin, and 
$P_j(\Phi_{\hi,j}|\bm{\theta})$ is the likelihood of $\Phi_{\hi,j}$ 
given $\bm{\theta}$:
\begin{equation}
		P_j(\Phi_{\hi,j}|\bm{\theta}) = 
		\frac{1}{\sqrt[]{2\pi}\Delta_j} 
		\exp \left\{ -\frac{\left[\Phi_{\hi,j} - \Phi^\prime_{\hi,j}(\bm{\theta})\right]^2}{2\Delta_j^2} \right\} 
\end{equation}
Here $\Phi^\prime_{\hi,j}(\bm{\theta})$ is the predicted HIMF in the $j$-th mass bin
derived by applying our HI estimator to {\tt SampleS} using the $\rm 1/V_{max}$ 
weighting scheme (see \autoref{sec:HIMF} for more details), and $\Delta_j$ is the error of 
$\Phi_{\hi,j}$.

\begin{figure*}[ht!]
	\centering
	\includegraphics[width=1.0\textwidth]{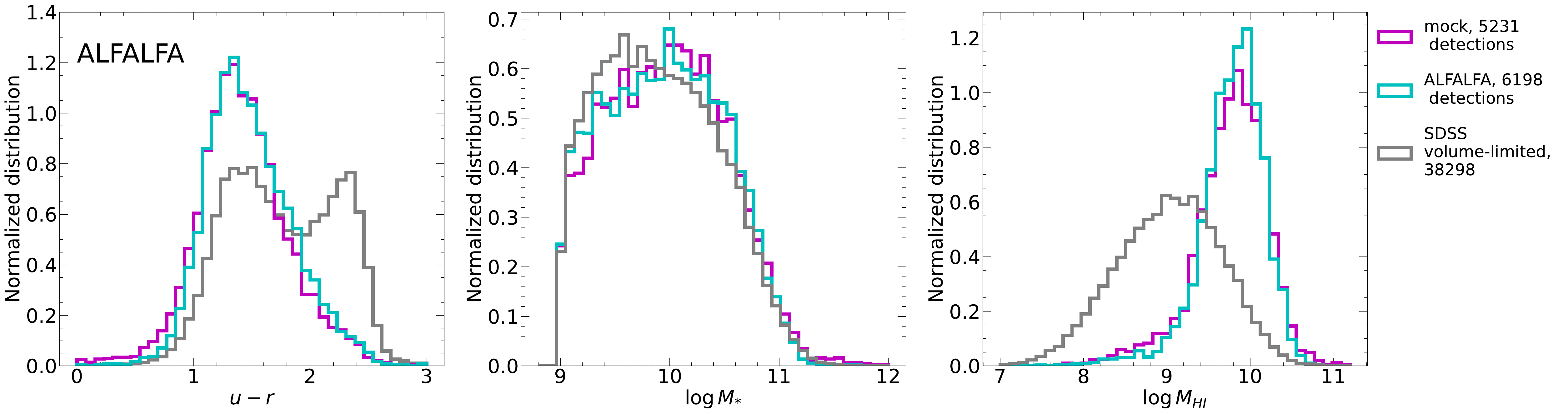}
	\includegraphics[width=1.0\textwidth]{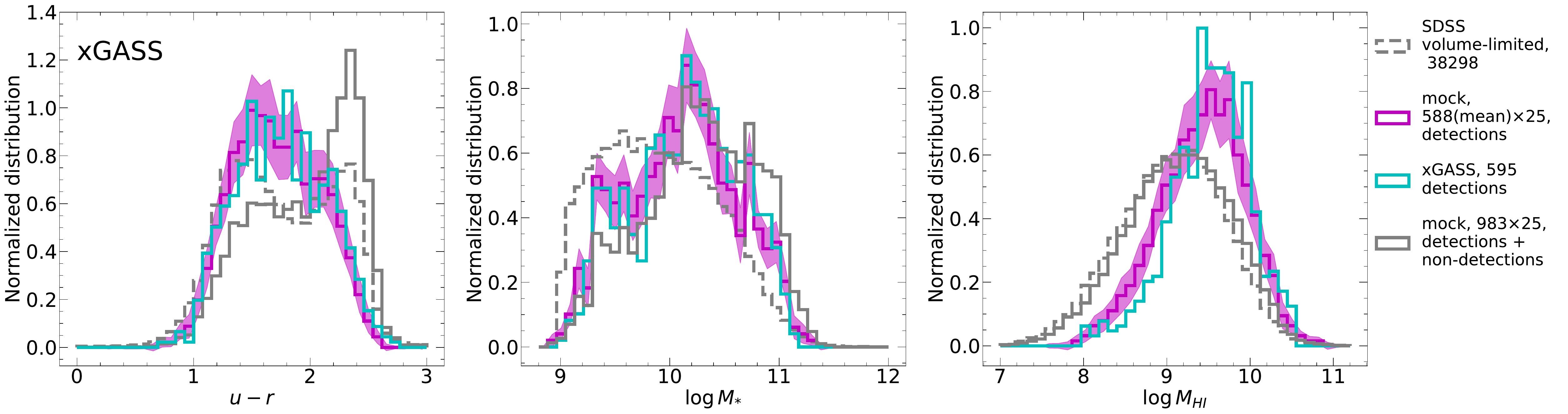}
	\caption{Normalized distribution of $u-r$, $\log M_*$ and $\log M_{\rm {\hi}}$ of galaxies.
	The cyan histogram represents the results of real samples and the magenta histogram of mock samples.
	Both the grey solid histogram in the top panel and the grey dashed histogram in the bottom panel represent the SDSS volume-limited sample used to construct ALFALFA mock catalog.
	The grey solid histogram in the bottom panel represents the mean distribution of the 25 mock xGASS samples.
	The magenta shaded area shows the $1-\sigma$ scatter of the 25 mocks.}
	\label{fig:mock_HI_survey}
\end{figure*}

\begin{figure*}[ht!]
	\centering
	\includegraphics[width=0.75\textwidth]{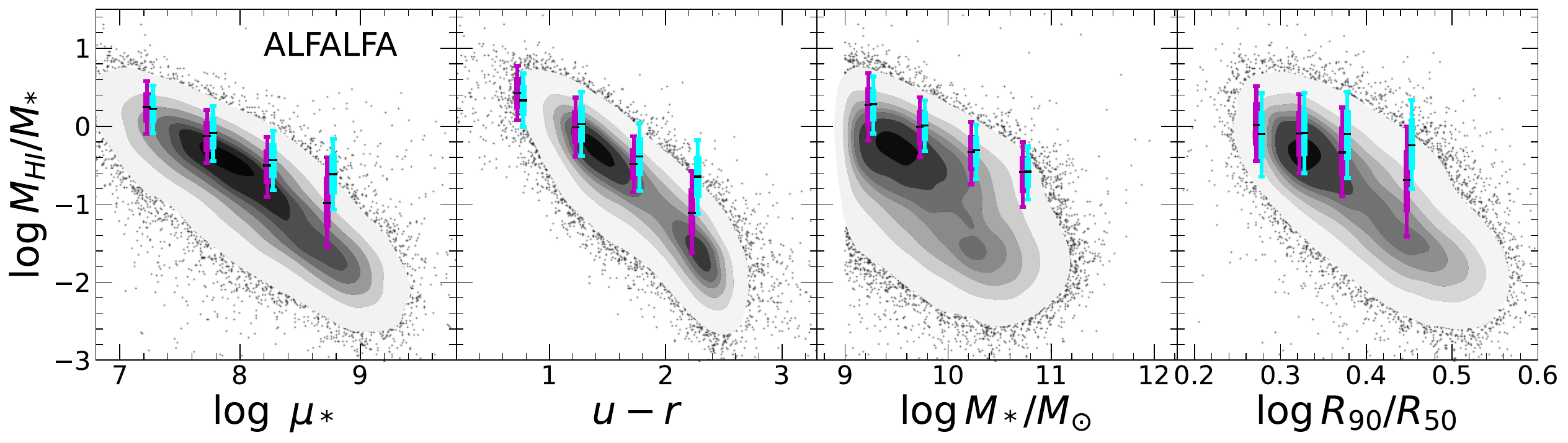}
	\includegraphics[width=0.75\textwidth]{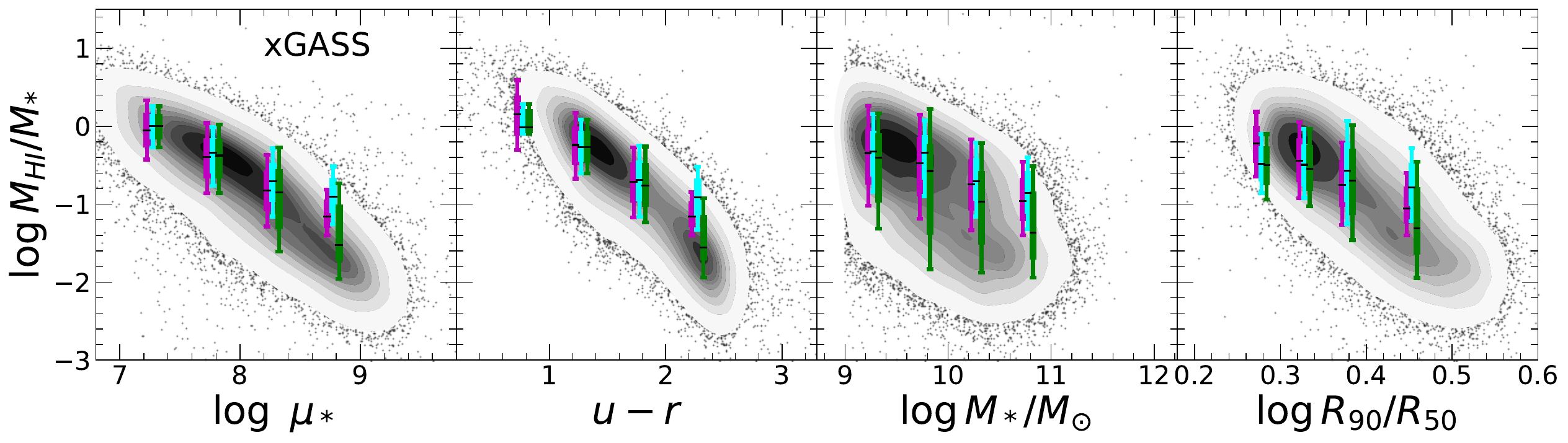}
	\caption{\hi\ mass fraction as a function of galaxy properties for ALFALFA(top)/xGASS(bottom) mock and real survey. 
		From left to right, the horizontal axis represents galaxy surface mass density,
		$u-r$ color index, stellar mass and galaxy concentration respectively.
		The y-axis is the observed {\hi} mass fraction for {\hi}-detected galaxies in real surveys
        and the estimated {\hi} mass fraction for galaxies in mock surveys.
        The grey contours represent the volume-limited sample used in the mock ALFALFA survey.
        The contours include 5\%, 15\%, 25\%, ..., 95\% of the total sample.
        Galaxies outside these contours are shown as grey dots.
		The boxes show the $25\%-75\%$ percentile range of galaxies detected in mock surveys (magenta) or real surveys (cyan) in each bin.
        The vertical error bars show the $5\%-95\%$ percentile range.
        The green boxes show the distribution of the total xGASS sample ({\tt SampleX}, detections + non-detections)}
	\label{fig:mock_HI_survey_2}
\end{figure*}

The same MCMC sampling as used in the first step is applied here 
to derive $P(\bm{\theta}|\Phi_{\hi},\bm{D})$ in \autoref{eqn:model2}.
The results are plotted in \autoref{fig:posterior} as red contours.
To guarantee convergence we have checked the 
autocorrelation time $\tau$ and we find $N / \tau > 80$ 
(where $N$ is the length of Markov chains) for all parameters, 
indicating that the Markov chains have well converged.
As expected, $P(\bm{\theta}|\bm{D})$ and $P(\bm{\theta}|\Phi_{\hi},\bm{D})$ are consistent 
with each other and the latter are better constrained thanks to the additional information 
provided by the HIMF. In particular, the posterior distributions of $c$ and $e$ are 
narrowed significantly. We adopt as our best model with 
maximum posterior probability: 
$a=-0.42\pm 0.20,\ b=-0.82\pm 0.19,\ c=0.15\pm 0.11,\ d=-2.57\pm 1.30,\ 
q=3.57\pm 0.95,\ c_a=0.10\pm 0.08$ and $c_b= -1.20\pm 0.76$. 
The average \hi-to-stellar mass ratio of galaxies at fixed stellar mass 
is predicted to have negative correlations with $\mu_\ast$, 
$u-r$ and $R_{90}/R_{50}$, which is consistent with direct observations
\citep[e.g.][]{2004ApJ...611L..89K,2009MNRAS.397.1243Z,2010MNRAS.403..683C}.
There is some degeneracy among model parameters. For instance, 
$q$ shows negative correlations with $a$ and a positive correlation 
with $b$. We note that $c_a$ and $c_b$ are strongly 
degenerated, possibly due to the strong constraint to the scatter 
from the data.

\subsection{Tests with mock {\hi} catalogs} 
\label{sec:mock_HI_survey}

Our \hi\ mass estimator is not simply a predictor of the mean value 
for galaxies of a given set of optical properties (as usually the case in 
previous studies), but rather it includes the variance of individual galaxies. 
For an individual galaxy with real \hi\ measurement, the estimated 
\hi\ mass may be different from the real value because of the scatter.
Therefore, we cannot directly compare the estimated and real 
values of the \hi\ mass for individual galaxies in ALFALFA and xGASS.
Instead, we can test our {\hi} estimator only statistically, by comparing 
the predicted distribution of \hi\ mass with that obtained from real 
observations. To facilitate a fair comparison, we also need mock catalogs 
constructed to include the same selection effects as xGASS or ALFALFA. 
If our {\hi} estimator provides an unbiased \hi\ prediction for a sample of 
general-population galaxies (rather than individual galaxies) selected 
from an optical survey, the mock catalog should reproduce the 
\hi\ mass distribution obtained from the real sample. 

From the ALFALFA$\alpha$.100 sample, we select a 100\% complete 
sample of $6198$ \hi-detected galaxies with $z<0.05$ and $\lgmstar/\msun>9$, 
and in the sky area of $138^{\circ}<\alpha<232^{\circ}$ and  
$0^{\circ}<\delta<36.5^{\circ}$, the overlapping sky coverage of the 
SDSS and ALFALFA $\alpha .100$ footprints. We require the lower mass 
limit of $\lgmstar/\msun>9$ in order to keep consistency with the 
mass limit of the calibration sample {\tt SampleX}.
In addition, galaxy samples selected with $z<0.05$ and $\lgmstar/\msun>9$
are actually volume-limited, and so we do not need to worry about 
sample incompleteness when comparing our mock catalog with the ALFALFA 
sample. Using the ALFALFA 100\% completeness limit is for the same 
consideration. The 100\% completeness limit is 
obtained by extrapolating the $25\%,\ 50\%$ and $90\%$ completeness limits 
derived by \cite{Haynes_2011} using the distribution of ALFALFA 
extragalactic sources in the plane defined by the \hi\ integrated flux density
($S_{21}$) and the line profile width ($W_{50}$). 
To construct the corresponding mock catalog, we start with a 
volume-limited sample of SDSS galaxies selected from {\tt SampleS} 
that covers the same sky area, stellar mass and redshift ranges 
as the ALFALFA 100\% complete sample. This volume-limited sample 
contains $38,298$ galaxies. For each galaxy we then apply our 
$\log(\mhi/\mstar)$ estimator to obtain an \hi\ mass, and 
calculate $S_{21}$ according to the estimated \hi\ mass and 
the redshift of the galaxy. We estimate the line width $W_{50}$ for 
the galaxy using the stellar mass Tully-Fisher relation given in 
\citet{10.1093/mnras/stx1701} with inclination considered. 
Considering that the \hi\ masses in the ALFALFA 
catalog are not corrected for \hi\ self-absorption and for a 
fair comparison between our mock catalog and the ALFALFA sample,
we have estimated an \hi\ mass for the mock galaxy without 
applying the self-absorption correction. The galaxy 
is included in the mock catalog if its \hi\ flux density is above 
the 100\% detection limit corresponding to its flux density and 
line width. A total of $5231$ galaxies are selected into the mock catalog. 

For xGASS, we take all the 983 galaxies including 595 \hi-detected 
galaxies with $\mhi/\mstar$ above a detection limit of 4\%, and 388 
non-detections below the limit. The detection threshold of the \hi-to-stellar 
mass ratio is chosen to be slightly higher than the usually quoted 
limit of $\sim1.5\%$ to ensure that the subsample of detections is 
truly complete down to the limit. We form a set of 25 mock catalogs, 
each constructed by randomly selecting 983 galaxies from {\tt SampleS} 
in the same stellar mass and redshift ranges either as the GASS survey
($10.2<\log \mstar/\msun<11.5$ and $0.025 < z < 0.05$) or 
as the extended xGASS survey ($9.0<\log \mstar/\msun<10.2$ and 
$0.01 < z < 0.02$). In addition, we require each of the mock catalogs 
to follow the same stellar mass distribution as the xGASS sample. 
The mock galaxies with $\mhi/\mstar>4\%$ are taken as detections 
and others are non-detections. On average, the mock catalogs contain 
588 detections, very close to the real sample (595). 

In \autoref{fig:mock_HI_survey} we compare the distribution of 
$u-r$, \lgmstar\ and \lgmhi\ for the mock and real samples, with the 
upper and lower panels for ALFALFA and xGASS, respectively. 
The mock catalogs well reproduce the 
distribution of \mhi\ for \hi-detected galaxies in both surveys, 
indicating that our estimator is able to provide unbiased
\hi\ estimates for the general population of galaxies. 

The mock catalog of ALFALFA also reproduces the 
distributions of $u-r$ and \lgmstar\ in the real sample. The distribution 
of the parent SDSS volume-limited sample used to construct the ALFALFA 
mock catalog is plotted in the upper panels for comparison. 
As expected, the ALFALFA detections are biased to relatively blue and 
\hi-rich galaxies. The full sample of xGASS
including both detections and non-detections show a similar distribution
in $u-r$ (although with a slightly higher fraction of blue galaxies) and 
the same distribution in \mhi\ when compared to the SDSS volume-limited 
sample. This indicates that xGASS is a more representative sample 
of the general population, and thus more appropriate for calibrating 
the \hi\ mass estimator, than ALFALFA. 

In \autoref{fig:mock_HI_survey_2} we plot the 
\hi-to-stellar mass ratio as functions of $\mu_\ast$, $u-r$, \mstar$/$\msun\ 
and $R_{90}/R_{50}$ for the real and mock catalogs. For the  mock 
catalogs and the SDSS volume-limited sample the \hi-to-stellar mass 
ratios are the values predicted by our estimator, while the observed values 
are plotted for real galaxies in ALFALFA and xGASS.
The mock catalogs of \hi-detected galaxies agree well 
with real \hi\ samples for both surveys. When compared to the model 
predictions based on the SDSS volume-limited sample, the
samples of detection from both surveys and the corresponding mock catalogs 
are biased for gas-rich galaxies, and this bias becomes more significant for 
galaxies with larger stellar mass, redder color, higher surface density, and more 
concentrated light distribution. When including both detections and 
non-detections (the green boxes), the xGASS mock catalogs closely follow the trend of the 
volume-limited sample, suggesting again that the xGASS survey represents well 
the general population of galaxies.

\section{{H\sc{i}} gas contents in galaxies and in dark matter halos} \label{sec:results}

\subsection{The {\rm {\hi}} Mass Function of galaxies} 
\label{sec:HIMF}

Using the constrained \hi\ mass estimator given above, we can predict the 
\hi\ mass function (HIMF) starting from an optical sample. For this purpose we use 
the SDSS galaxy sample ({\tt SampleS}) for which we have estimated an \hi\ mass 
for each galaxy using the best model parameters and their dispersion.
We use two different methods to estimate the HIMF, as detailed below.

In the first method, we use the $1/V_{\rm max}$-weighting scheme
following \cite{Li-White-2009} and \cite{2019ApJ...872..180C} in their 
estimates of the stellar mass function of galaxies. For each galaxy $i$ 
we determine $V_{\textrm{max},i}$, the total comoving volume of the survey 
out to the maximum redshift $z_{\textrm{max},i}$ at which the galaxy in 
question would meet the apparent magnitude criterion of our sample $r\leq 17.6$. 
Luminosity evolution and $K$-correction are included in the calculation of 
$z_{\textrm{max},i}$. The HIMF is then estimated as 
\begin{equation}\label{eqn:HIMF}
    \Phi(\mhi)\Delta \log \mhi = 
    \sum_i (f_{\textrm{sp},i} V_{\textrm{max},i})^{-1} \frac{\rho_{\rm u}}{\rho(V_{\textrm{max},i})},
\end{equation}
where the sum runs over all galaxies with \hi\ mass in the range 
$\mhi\pm 0.5\Delta \log \mhi$. Here $f_{\textrm{sp},i}$ is the spectroscopic 
completeness which varies across the survey area and is defined as the 
fraction of photometrically-selected targets that are spectroscopically 
observed and included in our sample. The last term in the equation 
$\rho_{\rm u}/\rho(V_{\textrm{max},i})$ is the ratio of the average 
mass density within the whole survey volume to that within $V_{\textrm{max},i}$,
as determined in \citet{2019ApJ...872..180C} based on the ELUCID 
simulation \citep{2014ApJ...794...94W}, a constrained simulation in the 
SDSS volume,  in order to account for the effect of cosmic variance caused by 
the limited sample volume at low redshift.

In the second method, we first estimate the $r$-band luminosity function 
(LF) from {\tt SampleS}, $\phi(M_r)$, using the same 
$1/V_{\rm max}$-weighting scheme as described above. We then divide 
all galaxies in {\tt SampleS} into a successive sequence of non-overlapping 
$M_r$ bins, $M_{r,k}\leq M_r<M_{r,k+1}~(k=1,...,N_m)$, 
with a fixed logarithmic interval of $\log(M_{r,k+1}/M_{r,k})=0.2$ dex. For the 
$k$-th interval we obtain the \hi\ mass distribution of sample galaxies 
$\Psi_k(\mhi\pm0.5\Delta\mhi)$, normalized so as to have
$\sum \Psi_k\Delta\mhi = 1$. The HIMF of the whole sample can then be estimated as 
\begin{equation}\label{eqn:method2}
        \Phi(M_{\hi}) = \sum _{k=1} ^{N_m} \left[
        \Psi _k(M_{\hi})\int_{M_{r,k}}^{M_{r,k+1}}\phi(M_r)dM_r\right],
\end{equation}
where the sum runs over all the absolute magnitude bins for a given \hi\ mass 
range $\mhi\pm 0.5\Delta\mhi$. 
In principle the two methods should lead to identical results. The advantage 
of the second one is that the complex selection effects of the 
galaxy sample are already accounted for when estimating the LF, 
so that one can obtain the HIMF simply by counting the 
sample galaxies in bins of \mhi. This method is valid only when the sample 
galaxies at a given luminosity are not biased in their \hi\ mass distribution.
This should be true for an optically-selected sample like the SDSS 
that does not involve any \hi-related selections. We use the LF estimated 
from {\tt SampleS} by ourselves, not only for self-consistency but also 
to take advantage of the $1/V_{\textrm{max}}$ weights given in
\citet{2019ApJ...872..180C} with corrections for the cosmic variance 
effect in the local universe. 

\begin{figure*}
	\centering
	\includegraphics[width=\textwidth]{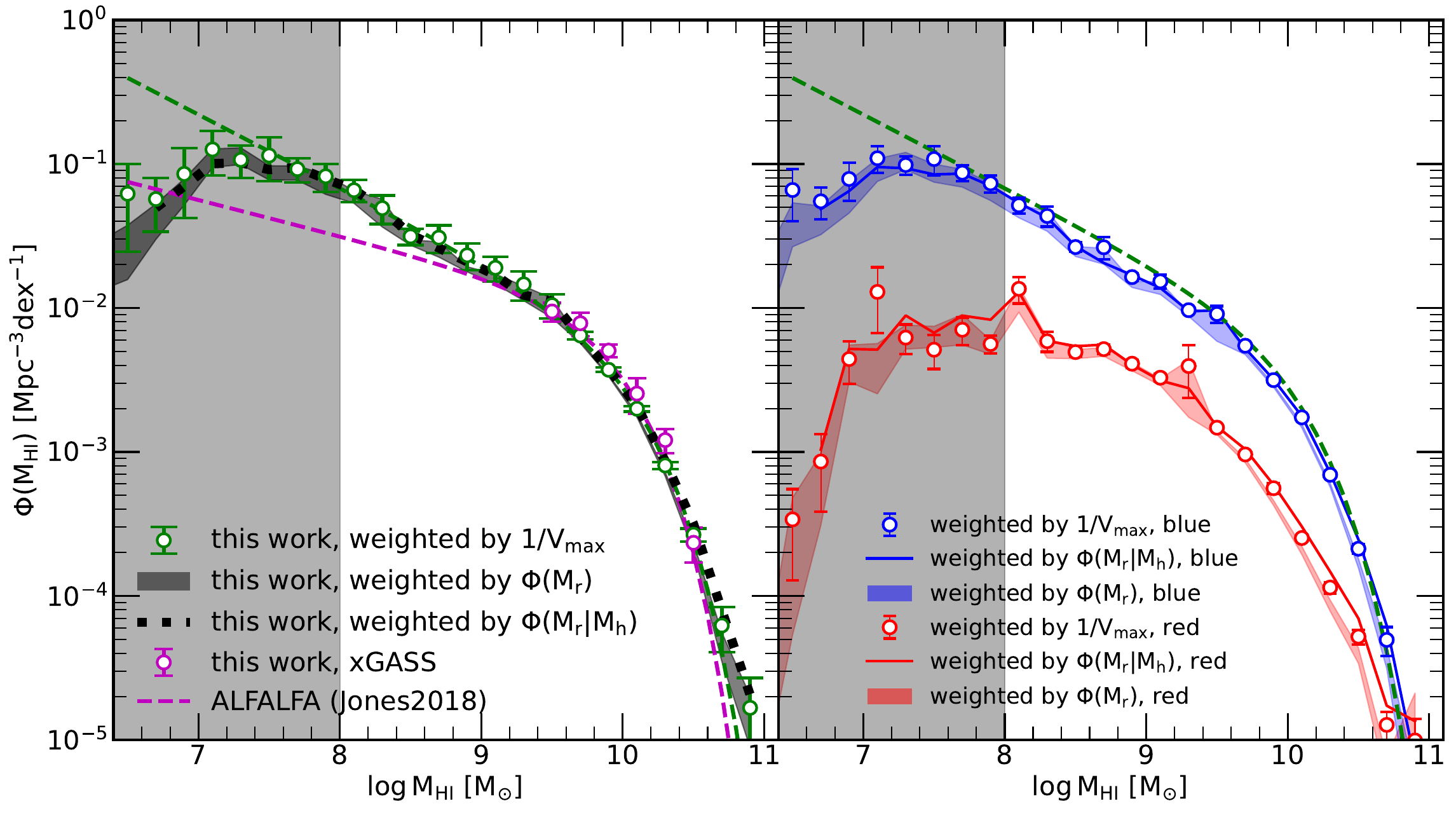}
	\caption{The {\hi} mass function of galaxies. Left panel: The \hi\ mass function of 
	all galaxies in the local universe. The green open circles with errorbars are the results
	of this work using $\rm 1/V_{max}$ weighting method. The green dashed line is the fit to
	the green circles. The dark shaded region shows the HIMF derived from our luminosity function.
	The thick dark dashed line is the HIMF derived from conditional luminosity function (see \autoref{subsec:discussion_CHIMF}).
	The magenta dashed line and open circles are the results of ALFALFA and xGASS (see \autoref{sec:bayes_likelihood}).
	The grey shaded region indicates the \hi\ mass regime where \texttt{SampleS} is incomplete and so the HIMF (green symbols) can only be 
	considered as lower limits of the true HIMF.
	Right panel: the \hi\ mass function of red and blue galaxies. The open circles are
	derived from $\rm 1/V_{max}$ method. The colored areas are derived from luminosity function.
	The solid lines are derived from conditional luminosity function.}	
	\label{fig:HIMF}
\end{figure*}

\begin{deluxetable}{llll}[ht!]
	\tablenum{1}
	\tablecaption{Schechter function parameters of CHIMFs\label{tab:Schechter_fits}}
	\tablewidth{0pt}
	\tablehead{
		\colhead{$\log (M_{200}/M_{\odot})$} & \colhead{$\Phi^\ast_{\textsc{Hi}}$} &  \colhead{$\log M^\ast_\textsc{Hi}$} & \colhead{$\alpha$}
	}
	\startdata
	\multicolumn{4}{l}{All galaxies:} \\ 
    $[12.00, 12.34)$ & $0.34\pm 0.04$ & $10.13\pm 0.04$ & $-1.17\pm 0.05$ \\ 
    $[12.34, 12.68)$ & $0.26\pm 0.02$ & $10.29\pm 0.03$ & $-1.41\pm 0.02$ \\ 
    $[12.68, 13.03)$ & $0.31\pm 0.02$ & $10.34\pm 0.02$ & $-1.54\pm 0.02$ \\ 
    $[13.03, 13.37)$ & $0.92\pm 0.11$ & $10.14\pm 0.04$ & $-1.42\pm 0.03$ \\ 
    $[13.37, 13.71)$ & $2.55\pm 0.24$ & $9.99\pm 0.04$ & $-1.23\pm 0.03$ \\ 
    $[13.71, 14.05)$ & $2.79\pm 0.32$ & $10.15\pm 0.03$ & $-1.50\pm 0.04$ \\ 
    $[14.05, 14.39)$ & $8.66\pm 1.10$ & $9.99\pm 0.04$ & $-1.41\pm 0.03$ \\ 
    $[14.39, 14.73)$ & $8.85\pm 1.09$ & $10.09\pm 0.04$ & $-1.56\pm 0.03$ \\ 
    $[14.73, 15.08)$ & $26.99\pm 3.15$ & $10.09\pm 0.03$ & $-1.53\pm 0.03$ \\  \hline
\multicolumn{4}{l}{Red galaxies:} \\ 
$[12.00, 12.34)$ & $0.10\pm 0.02$ & $9.84\pm 0.06$ & $-1.45\pm 0.06$ \\ 
$[12.34, 12.68)$ & $0.17\pm 0.03$ & $10.01\pm 0.05$ & $-1.43\pm 0.05$ \\ 
$[12.68, 13.03)$ & $0.16\pm 0.02$ & $10.22\pm 0.03$ & $-1.59\pm 0.03$ \\ 
$[13.03, 13.37)$ & $0.31\pm 0.06$ & $10.19\pm 0.06$ & $-1.51\pm 0.04$ \\ 
$[13.37, 13.71)$ & $0.89\pm 0.13$ & $10.00\pm 0.06$ & $-1.28\pm 0.04$ \\ 
$[13.71, 14.05)$ & $0.55\pm 0.09$ & $10.29\pm 0.05$ & $-1.68\pm 0.04$ \\ 
$[14.05, 14.39)$ & $3.75\pm 0.81$ & $9.97\pm 0.06$ & $-1.46\pm 0.05$ \\ 
$[14.39, 14.73)$ & $3.56\pm 0.71$ & $10.11\pm 0.06$ & $-1.69\pm 0.05$ \\ 
$[14.73, 15.08)$ & $9.02\pm 2.05$ & $10.16\pm 0.06$ & $-1.71\pm 0.05$ \\  \hline
%
\multicolumn{4}{l}{Blue galaxies:} \\ 
$[12.00, 12.34)$ & $0.30\pm 0.05$ & $10.13\pm 0.06$ & $-1.02\pm 0.07$ \\ 
$[12.34, 12.68)$ & $0.22\pm 0.03$ & $10.21\pm 0.05$ & $-1.26\pm 0.05$ \\ 
$[12.68, 13.03)$ & $0.19\pm 0.01$ & $10.36\pm 0.02$ & $-1.45\pm 0.01$ \\ 
$[13.03, 13.37)$ & $0.66\pm 0.05$ & $10.09\pm 0.03$ & $-1.34\pm 0.02$ \\ 
$[13.37, 13.71)$ & $1.72\pm 0.12$ & $9.97\pm 0.02$ & $-1.19\pm 0.02$ \\ 
$[13.71, 14.05)$ & $2.76\pm 0.22$ & $10.03\pm 0.03$ & $-1.30\pm 0.03$ \\ 
$[14.05, 14.39)$ & $5.23\pm 0.40$ & $9.98\pm 0.02$ & $-1.35\pm 0.03$ \\ 
$[14.39, 14.73)$ & $6.33\pm 0.49$ & $10.03\pm 0.03$ & $-1.36\pm 0.02$ \\ 
$[14.73, 15.08)$ & $23.48\pm 1.42$ & $9.96\pm 0.02$ & $-1.21\pm 0.02$ \\  \hline
\multicolumn{4}{l}{Satellite galaxies:} \\ 
$[12.00, 12.34)$ & $0.06\pm 0.01$ & $10.02\pm 0.02$ & $-1.65\pm 0.03$ \\ 
$[12.34, 12.68)$ & $0.14\pm 0.02$ & $10.04\pm 0.03$ & $-1.63\pm 0.03$ \\ 
$[12.68, 13.03)$ & $0.24\pm 0.02$ & $10.10\pm 0.03$ & $-1.67\pm 0.03$ \\ 
$[13.03, 13.37)$ & $0.79\pm 0.08$ & $10.04\pm 0.03$ & $-1.49\pm 0.03$ \\ 
$[13.37, 13.71)$ & $2.23\pm 0.18$ & $9.95\pm 0.03$ & $-1.27\pm 0.03$ \\ 
$[13.71, 14.05)$ & $2.82\pm 0.27$ & $10.09\pm 0.03$ & $-1.51\pm 0.04$ \\ 
$[14.05, 14.39)$ & $8.37\pm 1.07$ & $9.97\pm 0.04$ & $-1.42\pm 0.04$ \\ 
$[14.39, 14.73)$ & $8.58\pm 1.00$ & $10.08\pm 0.03$ & $-1.58\pm 0.03$ \\ 
$[14.73, 15.08)$ & $27.34\pm 3.05$ & $10.08\pm 0.03$ & $-1.53\pm 0.03$ \\ 
\enddata
\tablecomments{$\Phi^\ast_\textsc{Hi}$, $M^\ast_\textsc{Hi}$, and $\alpha$ 
are the characteristic amplitude, characteristic mass and the low-mass end slope 
of a Schechter function. In the last two halo bins, there is nearly no blue central galaxies
so no fitting parameters are given.}
\end{deluxetable}

\begin{deluxetable}{llll}[ht!]
\tablenum{2}
\tablecaption{Gaussian function parameters of CHIMFs\label{tab:Gaussian_fits}}
\tablewidth{0pt}
\tablehead{
\colhead{$\log (M_{200}/M_{\odot})$} & \colhead{$\Phi^\ast_\textsc{Hi}$} & \colhead{$\sigma$} &
\colhead{$\log M^\ast_\textsc{Hi}$} 
}
\startdata
\multicolumn{4}{l}{Central galaxies:} \\ 
$[12.00, 12.34)$ & $0.606\pm 0.017$ & $0.479\pm 0.010$ & $8.877\pm 0.013$ \\ 
                 & $0.369\pm 0.024$ & $0.391\pm 0.016$ & $9.714\pm 0.026$ \\ 
$[12.34, 12.68)$ & $0.742\pm 0.024$ & $0.460\pm 0.010$ & $9.072\pm 0.016$ \\ 
                 & $0.217\pm 0.021$ & $0.360\pm 0.020$ & $9.810\pm 0.037$ \\ 
$[12.68, 13.03)$ & $0.794\pm 0.032$ & $0.455\pm 0.009$ & $9.271\pm 0.021$ \\ 
                 & $0.165\pm 0.009$ & $0.442\pm 0.014$ & $9.932\pm 0.024$ \\ 
$[13.03, 13.37)$ & $0.851\pm 0.029$ & $0.444\pm 0.007$ & $9.371\pm 0.017$ \\ 
                 & $0.107\pm 0.009$ & $0.467\pm 0.016$ & $9.940\pm 0.041$ \\ 
$[13.37, 13.71)$ & $0.867\pm 0.024$ & $0.451\pm 0.006$ & $9.445\pm 0.014$ \\ 
                 & $0.099\pm 0.007$ & $0.443\pm 0.026$ & $10.013\pm 0.034$ \\ 
$[13.71, 14.05)$ & $0.895\pm 0.022$ & $0.457\pm 0.007$ & $9.559\pm 0.011$ \\ 
                 & $0.072\pm 0.010$ & $0.573\pm 0.024$ & $10.012\pm 0.069$ \\ 
$[14.05, 14.39)$ & $0.921\pm 0.037$ & $0.453\pm 0.009$ & $9.615\pm 0.016$ \\ 
                 & $0.016\pm 0.001$ & $0.419\pm 0.033$ & $10.051\pm 0.041$ \\ 
$[14.39, 14.73)$ & $0.930\pm 0.054$ & $0.412\pm 0.015$ & $9.686\pm 0.024$ \\ 
                 & $-$ & $-$ & $-$ \\ 
$[14.73, 15.08)$ & $0.948\pm 0.063$ & $0.425\pm 0.019$ & $9.719\pm 0.028$ \\ 
                 & $-$ & $-$ & $-$ \\ 
\enddata
\tablecomments{$\Phi^\ast_\textsc{Hi}$, $M^\ast_\textsc{Hi}$, and $\sigma$ 
are the amplitude, center and width of a Gaussian profile. For each halo bin 
the CHIMF of central galaxies are fitted with two Gaussians, corresponding to 
the two rows of parameters in the table (The first row for red and the second row for blue).}
\end{deluxetable}

In the left panel of Figure \ref{fig:HIMF}, we show the HIMF derived for our 
sample using the two methods described above, with green circles with error 
bars for the first method and the grey-shaded band for the second 
method. The errors of the HIMF from both 
methods are estimated using the bootstrap re-sampling method, which accounts for 
uncertainties due to sampling, but does not include the cosmic variance and
uncertainties in the \hi\ mass estimate. 
As expected, the HIMFs from the two methods are in good agreement 
with each other over the entire mass range. The HIMF increases as one 
goes from high to low \mhi\, but at $\mhi\sim10^8\msun$ the HIMF flattens 
and even decreases slightly towards the low-mass end. This behavior at the 
low-mass end is not real, but caused by the incompleteness of the SDSS 
sample at stellar masses below \mstar$\sim10^8$\msun, corresponding to 
a mean \hi\ mass of \mhi$\sim10^8$\msun\ according to our 
\hi\ mass estimator. Thus, our HIMF at $\mhi<10^8\msun$ should be taken 
as a `lower limit'. We fit our HIMF at $\mhi>10^8\msun$ with a single 
Schechter function \citep{Schechter-1976} and plot the result as the dashed 
green line in \autoref{fig:HIMF}. The parameters of the fit are 
$\Phi_{\hi}^\ast=2.61\pm 0.2 \times 10^{-3} {\rm Mpc^{-3}dex^{-1}}$, 
$\log M_{\textsc{hi}}^\ast/M_{\odot}=10.07\pm 0.02$, 
and $\alpha=-1.51\pm 0.02$. Integrating the best fit of our HIMF gives the
cosmic density of the \hi\ gas locked in local galaxies: 
$\Omega_{\rm {\hi}}=4.1\pm 0.1 \times 10^{-4}$, which is 17\% and 5\% higher 
than that from HIPASS \citep{2005MNRAS.359L..30Z} and 
ALFALFA \citep{2018MNRAS.477....2J}, respectively. 
For comparison we plot the Schechter function fit of the ALFALFA HIMF 
derived by \citet{2018MNRAS.477....2J} and the xGASS HIMF derived by 
us in \autoref{sec:bayes_likelihood}. 
We note that we have corrected the self-absorption effect for 
the estimated \hi\ masses in order for an HIMF estimate that is as close as 
possible to the true HIMF of the local Universe. Our tests showed that the 
self-absorption correction has a median value of $< 0.03$ dex, and so 
it should make little difference in the HIMF. Therefore, the comparison with 
the ALFALFA HIMF still makes sense, although the ALFALFA HIMF is based on 
uncorrected \hi\ masses. The HIMF estimated from 
the optical sample agrees well with both the ALFALFA and xGASS HIMFs
at \lgmhi$>9.5$ where we have all the HIMF estimates available.
At lower masses our HIMF differs from that of ALFALFA in that it has increasingly 
higher amplitudes and a steeper slope as the \hi\ mass decreases from 
$\mhi\sim 10^9\msun$ down to $\mhi\sim10^8\msun$. 
The difference is a factor of about 2 at $\mhi\sim10^8\msun$. 

In the right panel of Figure \ref{fig:HIMF} we show the HIMF derived 
in the same way, but separately for red and blue galaxies.  For this purpose 
we have classified each galaxy in {\tt SampleS} as red or blue according to 
the luminosity-dependent $u-r$ color demarcation given by \citet{2004ApJ...600..681B}.
As can be seen, the two methods provide very consistent results 
even when galaxies are divided into red and blue subsamples. 
The HIMF of the blue galaxy population is higher than 
that of the red population  over the entire $\mhi$ range except for the largest 
masses where the amplitudes of the two HIMFs become comparable. 
Clearly, the total HIMF of all galaxies is dominated by blue galaxies.

\begin{figure*}
	\plotone{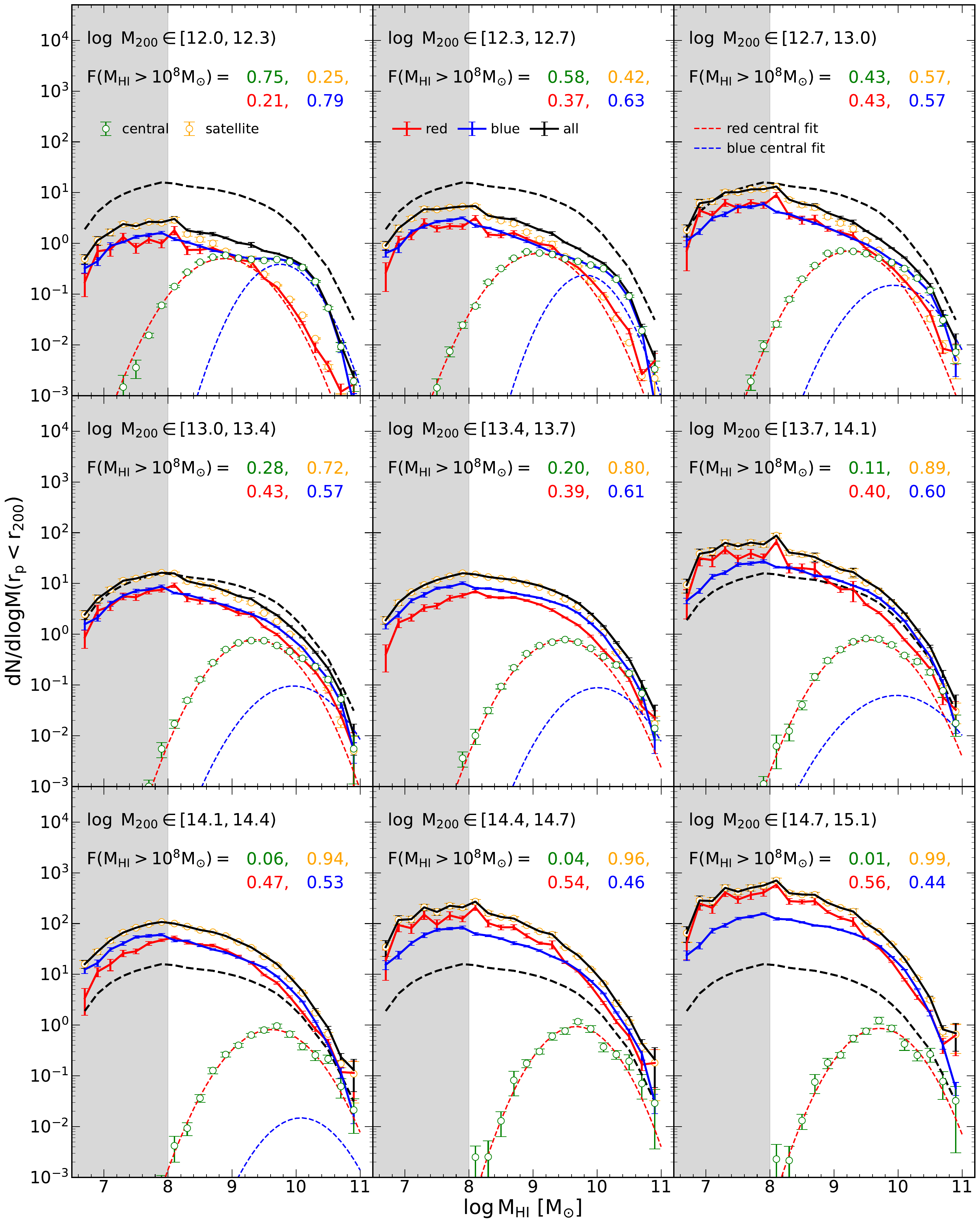}
	\caption{Conditional {H\sc{i}} mass function. The orange and green open circles represent
		the CHIMFs of satellites and centrals respectively. The black solid lines represent the CHIMFs of all galaxies.
		For comparison, the CHIMF of all galaxies in halos of $\log M_h\in [13.4,13.7)$ is plotted
		as the black dashed line in each panel.
		The red and blue solid lines are the CHIMFs of red and blue galaxies.
		The fitting result of central galaxies is shown as red and blue dashed lines
		for red and blue central galaxies.
		The colored numbers in each panel shows the contribution of central(green)/satellite(orange)/red(red)/blue(blue)
		galaxies to the total \hi\ mass, calculated using corresponding CHIMFs above $\rm M_{HI}>10^8 M_{\odot}$.
		The grey shaded region shows the {\hi} mass range
		where the CHIMFs suffer from incompleteness.}
	\label{fig:CHIMF_all}
\end{figure*}

\subsection{The conditional {\rm {\hi}} Mass Function} \label{sec:CHIMF}

The second statistic we can predict is the conditional \hi\ mass function 
(CHIMF), $\Phi(M_{\hi}|M_h)$, which is the HIMF of galaxies hosted by 
dark matter halos of a given mass $M_h$. A number of studies 
have previously attempted to obtain the HIMF of galaxies in groups/clusters
based on \hi\ observations of a small number of nearby groups 
\citep[e.g.][]{Freeland-2009,Kilborn-2009,Pisano2011}, or for individual systems 
in the local Universe such as the Leo I group \citep{Stierwalt-2009}, the Sculptor 
group \citep{Westmeier2017} and the Virgo cluster \citep{Jones2018a}. 
Based on the SDSS group catalog and the ALFALFA survey \citet{Jones2020} 
have recently obtained HIMFs for galaxy groups in three halo mass bins. 
Taking advantage of both the large and complete galaxy sample from the SDSS 
and the unbiased \hi\ masses estimated with our estimator, here we perform 
a first attempt to obtain the CHIMFs for narrow ranges of dark matter halo 
mass and down to substantially low \mhi, as well as for red/blue galaxies and
central/satellite galaxies separately.

We adopt the second method described in \autoref{sec:HIMF} to 
estimate the CHIMFs, except that we use the conditional luminosity function
(CLF; $\phi(M_r|M_h)$) instead of the total LF in \autoref{eqn:method2}. 
The CLFs of galaxies in the local Universe were estimated by 
\citet[][hereafter Lan16]{2016MNRAS.459.3998L} based on the SDSS 
galaxy group catalog \citep{2007ApJ...671..153Y} and the SDSS
photometric sample down to a $r$-band absolute magnitude of 
$M_r=-12$. Lan16 considered a number of halo mass bins
ranging from $M_h\sim10^{12}\msun$ up to $\sim10^{15}\msun$. 
For each of the bins they obtained the CLFs separately for red and blue 
galaxies (divided by a luminosity-dependent $u-r$ color demarcation), 
as well as for central and satellite galaxies (with the central/satellite 
classification provided by the SDSS group catalog). Correspondingly, 
we estimate the CHIMFs for red/blue and central/satellite galaxies 
separately, applying the same color cut and central/satellite classification 
as in Lan16 to divide our sample galaxies into red/blue or 
central/satellite subsamples. We divide galaxies of a given subsample 
into bins of $M_r$. The CHIMF for the subsample in each halo mass bin,
$\Phi(M_{\hi}|M_h)$, is then given by \autoref{eqn:method2}, where the 
LF is replaced by the corresponding CLF $\phi(M_r|M_h)$. 
We note that we have adopted a luminosity limit of $M_r=-14.4$ 
when estimating the CHIMFs, to avoid large uncertainties caused by the 
rather noisy distributions of \mhi\ at lower luminosities.

In \autoref{fig:CHIMF_all} we show the CHIMFs of red/blue and 
central/satellite galaxies as well as the total CHIMF, for different 
halo mass bins as indicated in each panel. The total CHIMF of the 
halo mass bin of $13.37\leq\log(\mhalo/\msun)<13.71$ 
is repeated as the dashed black line in each panel for reference. 
From the CHIMFs we calculate the fraction of the total \hi\ mass 
in dark halos of given mass as contributed by red, blue, central and satellite 
galaxies with \mhi$>10^8$\msun. These mass fractions are indicated 
in each panel with different colors for different types of galaxies. 
A Schechter function can well describe all the HIMFs except those for 
central galaxies where a double Gaussian profile provides a better 
description for low-mass halos and a single Gaussian profile 
works better for high-mass halos. When divided into red and blue 
subsamples, central 
galaxies at fixed halo mass can each be modeled by a single 
Gaussian, as is shown in the figure where red and blue dashed lines 
are the Gaussian fits of red and blue centrals, respectively.
The Schechter function parameters for the total sample and 
the subsamples of red, blue and satellite galaxies in each 
halo mass bin are listed in \autoref{tab:Schechter_fits}, 
while the parameters of the two Gaussian functions for 
sub-samples of central galaxies are listed in \autoref{tab:Gaussian_fits}. 

As can be seen from \autoref{fig:CHIMF_all}, the amplitude 
of the total HIMF increases monotonically with increasing halo mass, 
indicating a positive correlation of the total \hi\ mass of galaxies 
with the dark matter mass of their host halos. 
The low-mass slope of the total HIMF 
in Milky Way-like halos ($\mhalo \sim 10^{12} \msun$) is relatively 
flat with $\alpha=-1.17$, which is consistent with previous works
where $\alpha\lesssim -1.2$ was found for nearby groups 
hosted by dark halos of similar mass \citep[e.g.][]{Freeland-2009,Kilborn-2009,
Pisano2011,Westmeier2017,Jones2020}. As halo mass increases, the 
low-mass slope becomes steeper, reaching a slope index of $\alpha<-1.5$ 
at the highest halo mass (see \autoref{tab:Schechter_fits}). 
\citet{Jones2020} also attempted to measure the HIMFs for 
SDSS galaxy groups of $\mhalo>13h^{-1}\msun$ using the ALFALFA survey, 
finding flat slopes at the low-mass end similar to groups of lower masses. 
As those authors pointed out, however, their low-mass slope was well 
sampled only for groups at $\mhalo<10^{13}\msun$ and the measurements 
at higher masses were uncertain due to the limited depth of the ALFALFA survey.
In addition, from \autoref{tab:Schechter_fits} we find the 
characteristic mass $M^\ast_\textsc{Hi}$ of the Schechter function 
for the total HIMF to span a narrow range of 
$M^\ast_\textsc{Hi}\sim1-2\times10^{10}\msun$. This result is consistent 
with early studies which found weak dependence of the turnover mass 
of galaxy HIMFs on environmental density \citep[e.g.][]{Moorman-2014,Jones-2016}.

When galaxies are divided by color, we find that the red and 
blue populations present similar CHIMFs for halos of different masses,
in terms of both amplitude and shape, which results in a comparable 
\hi\ mass fraction for the two populations in halos of different masses.
Quantitatively, the \hi\ mass fraction in red (blue) galaxies 
increases (decreases) steadily from 21\% (79\%) at \mhalo$\sim10^{12}$\msun\ 
to 56\% (44\%) 
at \mhalo$\sim10^{15}$\msun. It has been well established that blue galaxies 
dominate the central galaxy population in intermediate-to-low mass halos, 
and red galaxies are either (massive) centrals in massive halos or 
(less-massive) satellites in halos of different 
masses \citep{2009ApJ...695..900Y,2016MNRAS.459.3998L}.
Thus, massive halos are expected to contain more red galaxies than blue galaxies.
The fact that red and blue galaxies have comparable CHIMFs at given 
halo mass can be understood from the color-dependent 
\hi-to-stellar mass ratio of galaxies, which is smaller for 
redder galaxies.

\begin{figure*}
	\plottwo{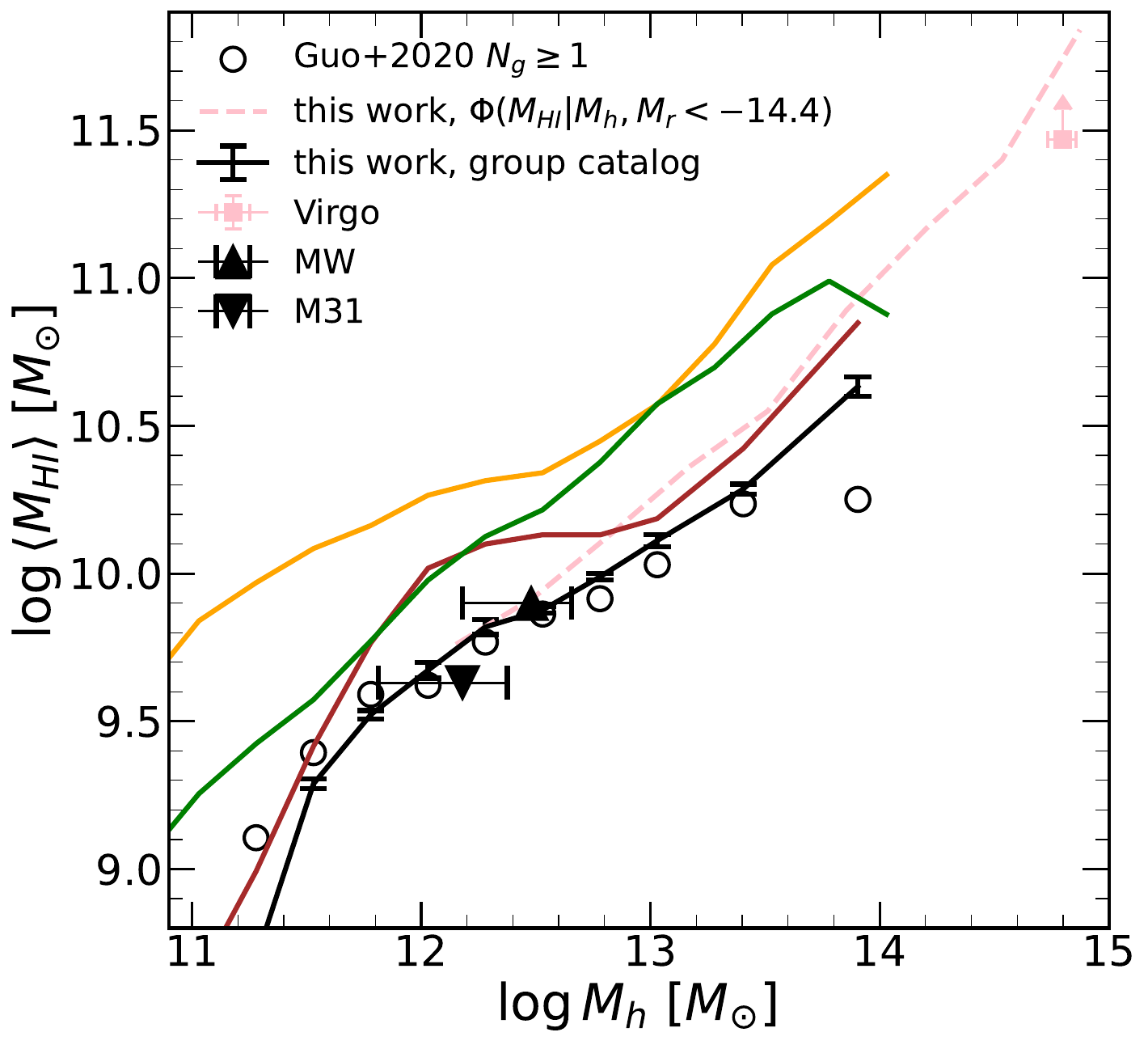}{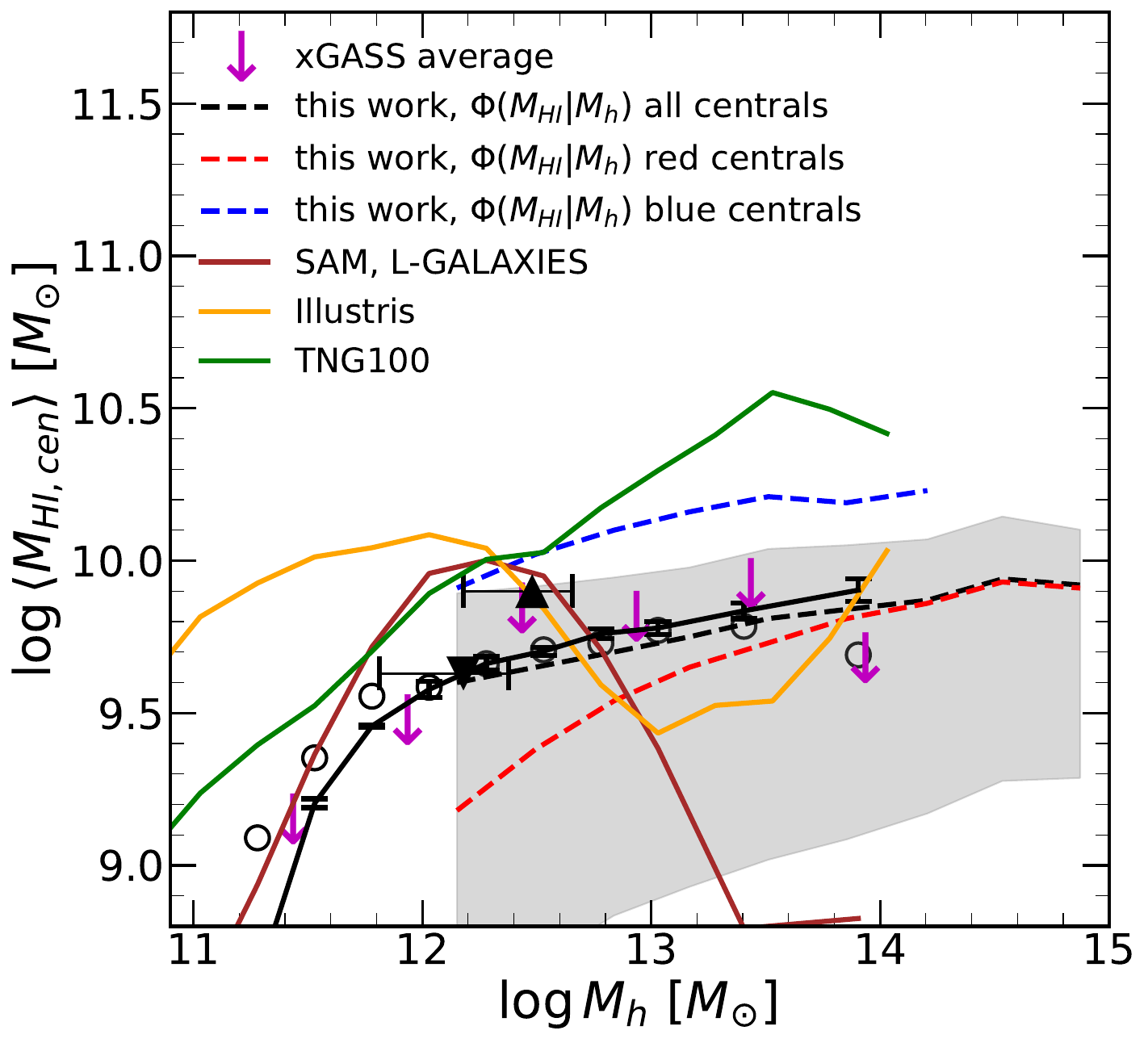}
	\caption{The left panel shows the average \hi\ mass of dark matter halos and
	the right panel shows the average \hi\ mass of central galaxies.
	In both panels, the black open circles are the result of \cite{2020ApJ...894...92G}
	in $N_g\ge 1$ case. For central galaxies, they have been corrected for the 
	confusion effect using our estimator.
	The black solid line is our result derived from 
	group catalog. The brown, orange and green solid lines represent 
	the result of L-GALAXIES semi-analytical model, Illustris simulation 
	and IllustrisTNG simulation, respectively.
	The black upward and downward triangles represent 
	the Milky Way and M31, respectively.
	In the left panel, the pink dashed line is the result of our CHIMF
	and the pink upward arrow with error bars 
	represents Virgo Cluster. 
	In the right panel, the black dashed line
	is the average central {H\sc{i}} mass derived from our CHIMFs, and the grey shaded
	region represents the 16\% to 84\% percentile of the central CHIMFs in each halo mass bin.
	The red and blue dashed lines are the mean \hi\ mass of red and blue central galaxies from our CHIMFs.
	The magenta downward arrows represent the upper limit of 
	\hi\ masses of central galaxies in xGASS sample.
}
	\label{fig:group_MHI_Mh}
\end{figure*}

For central galaxies, the CHIMFs appear to depend weakly on halo mass,
with the average \mhi\ increasing slightly and the width decreasing slightly 
with increasing halo mass. The two Gaussian components 
correspond to contributions by central galaxies of red and blue colors, 
respectively. Blue centrals contribute significantly to the total \hi\ of the central 
galaxy population only in low-mass halos, with a contribution that is 
comparable to that of red centrals in Milky Way-like halos and 
declines with halo mass. At \mhalo$\ga 10^{14}$\msun, 
the CHIMF of centrals is dominated by red galaxies, with a negligible
contribution from blue centrals.
Central galaxies as a whole contribute a significant amount of \hi\ mass 
only in low-mass halos, with a fraction of 75\% and 58\% 
in the two lowest-mass halo bins, respectively, and the fraction decreases rapidly 
at larger halo masses, becoming less than 10\% when the halo mass
exceeds $\mhalo\sim 10^{14}\msun$. In contrast to centrals, the satellite galaxy 
population has a Schechter CHIMF that varies with halo mass in a way similar 
to the total CHIMF. Consequently, the satellite population contributes 
a larger fraction of the total \hi\ mass as the halo mass increases. 
We would like to point out that, this result is not 
in conflict with the known fact that \hi-selected galaxy samples in the 
local Universe are dominated by field central galaxies \citep{Martin2012,Papastergis2013,Guo2017}. 
This is because our CHIMFs are limited to relatively massive halos with 
$\mhalo > 10^{12} \msun$, while field centrals are predominantly found 
in halos of lower masses. The total \hi\ budget of the galaxy population 
should be dominated by low-mass halos according to the power-law 
shape of the halo mass function.


\begin{figure*}
	\centering
	\plotone{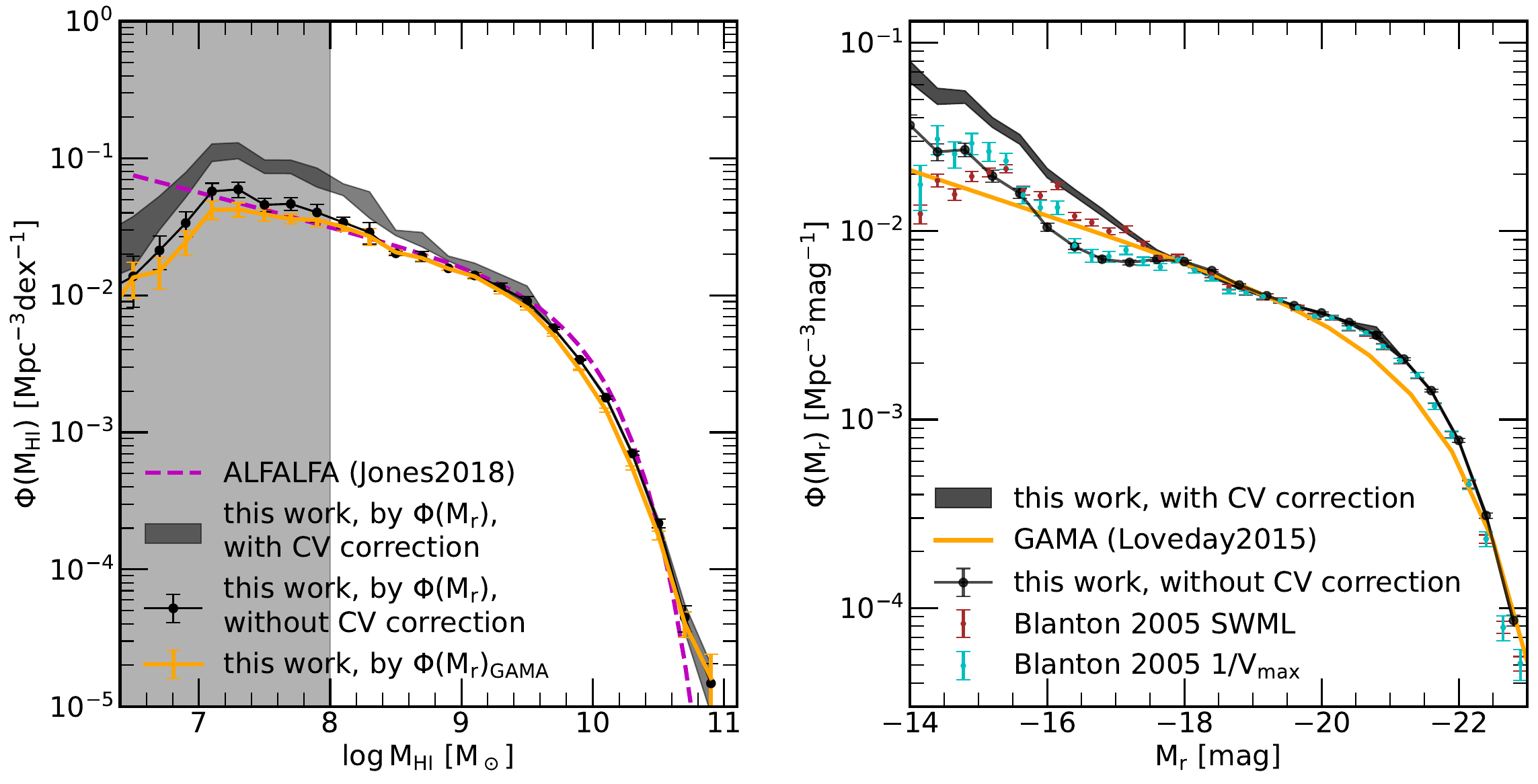}
	\caption{Left panel: the \hi\ mass function. The magenta dashed line is ALFALFA's result.
	The orange solid line is derived from GAMA LF.
	The black shaded region and solid line are derived from our LF with and without CV correction respectively.
	The grey shaded region shows the mass range suffering from sample incompleteness.
	Right panel: the orange solid line is the GAMA LF. The brown and cyan points with errorbars represent
	the LF of \cite[]{2005ApJ...631..208B} derived by SWML and $\rm 1/V_{max}$ method respectively.
	The dark region and points show our LF with and without CV correction.}
	\label{fig:GAMA}
\end{figure*}

\subsection{The \hi-halo mass relation} 
\label{sec:HI-halo mass relation}

One can obtain the total \hi\ mass locked in member galaxies in dark matter 
halos by integrating the CHIMFs  derived above. The result 
is shown as dashed lines in \autoref{fig:group_MHI_Mh}, where we plot the 
total \hi\ mass in all galaxies (left panel) and central galaxies (right panel) as 
a function of halo mass. The total \hi\ mass of all galaxies increases monotonically 
with halo mass, from \mhi$\sim8\times10^9$\msun\ for Milky Way-like halos with 
\mhalo$\sim10^{12}$\msun\  up to \mhi$\sim7\times10^{11}$\msun\ for Coma-like halos 
with \mhalo$\sim5\times10^{14}$\msun.  The total \hi\ mass in central galaxies shows 
a rather weak dependence on halo mass, with \mhi\ slightly increasing from 
$\sim4\times10^{9}$\msun\ to $\sim8\times10^{9}$\msun. 

The \hi-to-halo mass (HIHM) relation can also be estimated from a galaxy 
group catalog \citep{2017MNRAS.470.2982L} for which we have estimated 
the \hi\ mass for each galaxy using our estimator. To that end, we divide 
galaxy groups into a set of halo mass bins, and for groups in each bin we sum up the 
\hi\ mass of all member galaxies (or central galaxies) to obtain the HIHM
relation for all galaxies (or central galaxies). Each galaxy is weighted in the same way 
as in \autoref{eqn:HIMF} in order to correct for the sample incompleteness and 
cosmic variance effect. The relations derived this way are plotted as the solid black 
lines in \autoref{fig:group_MHI_Mh} over the range from \mhalo$\sim10^{11}$\msun to
\mhalo$\sim5\times10^{13} \msun$. The error bars are estimated from the scatter 
among 20 bootstrap samples.
For central galaxies, the HIHM relation derived from the group catalog 
is consistent with the average relation derived from the CHIMFs. 
For all galaxies as a whole, however, the relation derived from the group 
catalog agrees with the CHIMF-based result only at \mhalo$\sim1-2\times 10^{12}$\msun.
Above this mass, the group catalog gives lower \hi\ mass, and the difference 
increases with halo mass, reaching $\sim0.3$ dex at \mhalo$\sim5\times10^{13}$\msun.
Possible reasons for this difference will be discussed in the next section.

The HIHM relations estimated from the group catalog are in good agreement with 
those of \cite{2020ApJ...894...92G} obtained by stacking the ALFALFA datacubes, as shown
by open circles in \autoref{fig:group_MHI_Mh}. We have corrected the 
confusion effect for the average \mhi\ mass of central galaxies measured 
by \cite{2020ApJ...894...92G} by subtracting the contribution of satellite galaxies that 
reside near their central galaxy. Following \cite{2020ApJ...894...92G}, 
we use an angular aperture of 
$\textrm{max}$\{200kpc/$D_A,8^\prime$\} and a maximum velocity separation of 
300 km $s^{-1}$ to define the nearness. For comparison, the left panel of the same 
figure shows the total \hi\ mass for the Milky Way  
\citep[][]{2009ARA&A..47...27K}, M31 \citep[][]{2009ApJ...705.1395C} and the 
Virgo Cluster. The {\hi} mass of the Virgo Cluster is estimated by ourselves
by applying our {\hi} estimator to the Extended Virgo Cluster Catalog 
\citep[EVCC;][]{2014ApJS..215...22K}. We only use galaxies in EVCC that 
are brighter than $M_r=-14.4$, corresponding to the luminosity limit adopted above for 
our CHIMFs, and are classified to be member galaxies of the Virgo Cluster
based on the spherical symmetric infall model \citep{1994ApJ...422...46P}.
Thus, the resulting {\hi} mass should perhaps be considered as a lower limit of the 
total \hi\ mass in the cluster, as some {\hi} may be contained in galaxies 
that are not included in our estimate. 
Observationally, \cite{Jones2018a} measured the HIMF of Virgo Cluster using ALFALFA data. Integrating the Virgo HIMF of the "extended Virgo cluster sample" in their work and assuming a spherical volume with a radius of 3 Mpc gives an \hi\ mass of $\log \mhi(\rm{Virgo})/\msun \sim 11.14$, about $0.3$ dex lower than our result. However, we emphasize that since the definitions of Virgo galaxy sample used in these two works are quite different, this comparison is actually unfair. It is hard to tell whether our \hi\ estimation of Virgo Cluster is consistent with \cite{Jones2018a} or not.
Using central galaxies in \texttt{SampleX} we can also estimate the average 
HIHM relation, which is shown in the right-hand panel \ref{fig:group_MHI_Mh}. 
We have included both detections and non-detections in this estimation, and so the 
\hi\ masses from xGASS should be regarded as upper limits. As one can see, 
our HIHM relations are in broad agreement with all these previous measurements 
based on real observations of the \hi\ gas, indicating that our estimator 
is reliable also for member galaxies of groups. 


\begin{figure*}[ht!]
	\centering
	\plotone{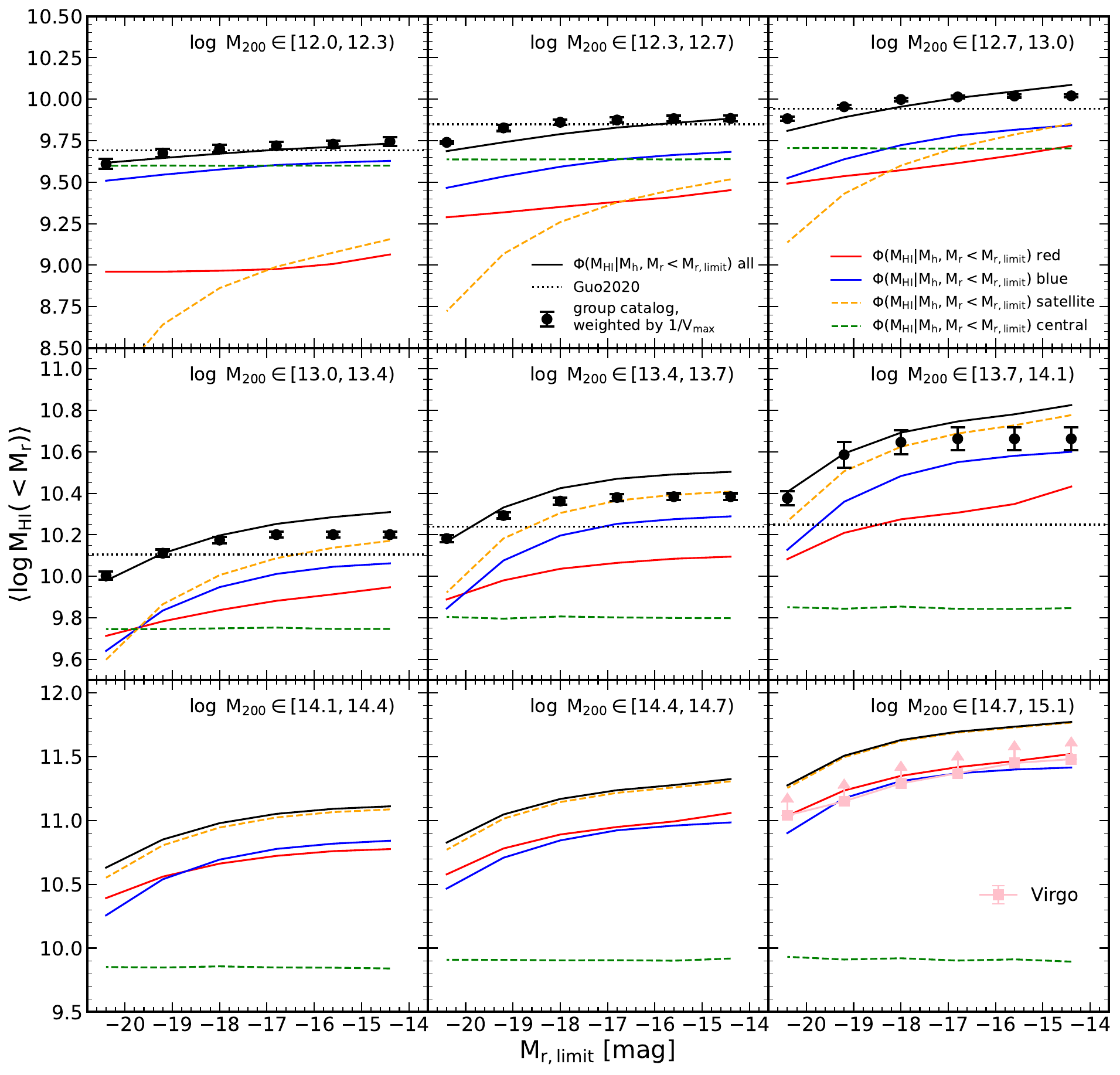}	
	\caption{The \hi\ mass in dark matter halos under different luminosity limits.
	In each panel, the black points represent the average \hi\ mass of all galaxies
	in halos derived using group catalog. The horizontal black dotted line shows the result of \cite[]{2020ApJ...894...92G}.
	The red/blue/green/orange lines represent the \hi\ mass in red/blue/central/satellite galaxies
	from our CHIMFs. The pink upward arrows show the result in the Virgo Cluster.}
	\label{fig:HIHM_limits}
\end{figure*}

\section{Discussion}
\label{sec:discussion}

\subsection[]{HIMF at the low mass end} \label{sec:discussion_HIMF}

\autoref{fig:HIMF} shows that the HIMFs predicted by our estimator are higher 
and steeper than the HIMF obtained from ALFALFA at masses below a few
$\times10^9$\msun. This discrepancy is predominantly caused by cosmic variance, 
which is corrected in our HIMFs but not in the ALFALFA result.
The cosmic variance is taken into account through the last term of \autoref{eqn:HIMF},
which was carefully determined by \cite{2019ApJ...872..180C} based on 
the ELUCID simulation, a constrained simulation with initial conditions
that are reconstructed to reproduce the dark matter density field in  
the SDSS volume \citep{2014ApJ...794...94W}. The local volume 
within $z\sim 0.03$ has a density that is lower than the mean density 
of the universe. Consequently, the SDSS under-represents faint galaxies 
in the universe because it can detect these galaxies only at the very 
low redshift. ALFALFA is also limited to very low redshift, and so 
is also affected by the cosmic variance.

The effect of cosmic variance is shown clearly in \autoref{fig:GAMA}.
In the right-hand panel,  we show the $r$-band luminosity functions (LFs), 
$\Phi(M_r)$, estimated from {\tt SampleS} with and without the cosmic variance 
correction, and compare them with the LFs obtained by \citet{Blanton_2005} using an 
early SDSS sample, and with the LF obtained by \citet{2015MNRAS.451.1540L} 
using the GAMA survey \citep{2011MNRAS.413..971D}. The latter two studies 
did not correct for the cosmic variance effect, and it is not surprising 
that their LFs are flatter and lower than our measurements at the faint end.
Our LF without including the cosmic variance correction shows a similarly 
lower amplitude at the low-mass end, confirming the importance of cosmic 
variance in the local universe. As pointed out in 
\cite{Blanton_2005}, the dip at $M_r\sim-17$ and the upturn at 
the faint end seen both in our uncorrected LF and the LF of \cite{Blanton_2005} 
(the one estimated with the $1/V_{\textrm{max}}$ method)
are produced by cosmic variance caused by the presence of 
large-scale structure in the sample volume, as the $1/V_{\textrm{max}}$
method is not capable of correcting such variance. The LF of \cite{Blanton_2005}
estimated using the step-wise maximum likelihood does not show such 
fluctuations at the faint end and is more consistent with the GAMA LF, 
but both are still lower than our LF corrected for the cosmic variance. 
In the left-hand panel of \autoref{fig:GAMA} we compare the HIMFs derived 
with \autoref{eqn:method2}, using our LFs with and without cosmic variance 
correction, as well as using the GAMA LF. The HIMFs without the correction are 
in good agreement with the ALFALFA result, demonstrating that the difference 
in the low-mass end between our corrected HIMF and the ALFALFA result 
is produced by the different treatments of the cosmic variance effect. 

\subsection{Conditional \hi\ mass functions} \label{subsec:discussion_CHIMF}

As demonstrated in \autoref{sec:mock_HI_survey}, our \hi-to-stellar mass 
estimator reproduces well the statistical properties of the mock catalogs 
that have the same selection effects as ALFALFA and xGASS. The 
application of our estimator to the SDSS sample also yields HIMFs that are in 
good agreement with the ALFALFA HIMF. All these indicate that our predicted CHIMFs 
are reliable. As a further test, we combine our CHIMFs for different halo masses 
to predict a total HIMF and compare it with the HIMFs estimated using other methods.
The black dashed line in the left panel of \autoref{fig:HIMF} shows the HIMF 
obtained from the sum of CHIMFs of different halo mass bins, each weighted 
by the number density of halos in the halo mass bin calculated using the halo mass 
function from \citet{2022MNRAS.515.2138D}. 
Since the CHIMFs can be estimated only down to Milky Way-size halos with 
\mhalo$\sim10^{12}$\msun\ due to the mass limit in the SDSS group catalog, 
we have included the contribution of halos of lower masses following the 
method described in \cite{2016MNRAS.459.3998L}. As can be seen from 
the figure, the HIMF estimated from the CHIMFs agrees very well with the 
HIMFs obtained with other methods. 

As we have shown, the Schechter function provides a good description 
for the CHIMFs of all/red/blue/satellite populations in halos of different 
masses, while a double Gaussian function is suitable for central galaxies 
(see \autoref{tab:Schechter_fits} and \autoref{tab:Gaussian_fits}). These 
results have important implications for models of galaxy formation. 
For instance, a recent study by \cite{2022arXiv220710414L} used 
the \hi-to-halo mass relation and clustering of {\hi} galaxies obtained 
by \cite{Guo2017} and \cite{2020ApJ...894...92G} to constrain 
empirical models of the \hi-halo relation by assuming that the \hi\ mass of 
galaxies depends only on dark matter halo mass plus a secondary halo property,
such as halo formation time and concentration.
Their best-fit model predicts CHIMFs that have a log-normal 
form, quite different from the Schechter function found here
(\autoref{fig:CHIMF_all}). This implies that the model must have missed 
some key ingredients that regulate the \hi\ gas content of galaxies.
Our \hi\ estimator involves four parameters, among which the stellar mass 
and color may be modeled by halo mass and formation time 
\citep[see][for a review]{2018ARA&A..56..435W}.
The other two parameters, surface stellar mass density and 
concentration, characterize the structure of galaxies and have been found 
to depend weakly on environments and properties of their host halos 
\citep[e.g.][]{2006MNRAS.368...21L,2009ARA&A..47..159B}.
On the other hand, the \hi\ mass fraction of galaxies may be stripped 
by environmental processes operating in dense regions,
such as tidal stripping and ram-pressure stripping
\citep[e.g.][]{1972ApJ...176....1G,1977Natur.266..501C,1982MNRAS.198.1007N,2006PASP..118..517B,2021ApJ...915...70W,2022ApJ...927...66W}.
These effects are expected to depend on both the stellar mass and 
surface mass density of galaxies \citep[e.g.][]{2012MNRAS.424.1471L,2013MNRAS.429.2191Z}. 
It is thus likely that physical processes related to both galaxy 
structure and environments may also play a role in determining 
\hi\ gas contents of galaxies. 

When divided according to color, red and blue central galaxies follow
a Gaussian CHIMF, rather than a Schechter function
(\autoref{fig:CHIMF_all}). This is similar to the conditional stellar mass 
(or luminosity) function of central galaxies found in previous studies 
\citep[e.g.][]{2016MNRAS.459.3998L, 2008ApJ...676..248Y}. However, as shown in 
\autoref{fig:group_MHI_Mh}, the average \hi\ mass of central galaxies 
is limited to a small range, \mhi$\sim4-8\times10^9$\msun\, for 
halos of Milky Way masses or larger.  The weak dependence on the halo mass 
holds even when central galaxies are divided into red and blue subsamples
(see \autoref{fig:group_MHI_Mh}). This result implies that, unlike the 
stellar content of central galaxies, which closely follows the growth of 
their host dark matter halos, the gas contents and gas-related 
evolution of central galaxies may be driven by internal and local processes. 
In massive halos, the central galaxies are predominantly red. 
Considering the large stellar mass and relatively small star formation 
rate as inferred by the red color, it is perhaps not surprising
to see the nearly constant \hi\ mass of these galaxies. 
However, the weak dependence on halo mass itself suggests that their 
gas-related quenching are unlikely regulated by the host halos, but 
rather some internal processes may be able to shut off star formation 
in these galaxies, making them red
but do not reduce their \hi\ gas contents significantly.

The total \hi\ mass of galaxies as a function of dark matter 
halo mass, obtained using our CHIMFs and the SDSS galaxy group catalog, 
should be understood as the total \hi\ mass locked in {\em optically-selected 
galaxies down to a certain limiting luminosity}. Note that the 
optical samples used to obtain the CLFs and to construct the group catalog 
have different depths. The group catalog of \citet{2007ApJ...671..153Y}
was constructed using the SDSS spectroscopic galaxy sample limited to $M_r\sim-19.5$.  
On the other hand, the CLFs of \citet{2016MNRAS.459.3998L} 
used the same group catalog to 
identify groups/halos, but with member galaxies identified from a much deeper 
photometric sample. This explains the increasing difference in the 
total \hi\ mass between the prediction of the CLFs and that of 
the group catalog, as seen in \autoref{fig:group_MHI_Mh}. The difference 
is larger for more massive halos where groups are observed to 
higher redshifts and thus may miss more satellite galaxies 
below the luminosity limit. To better understand this effect, we have repeated 
the estimation of the \hi-to-halo mass relation using different luminosity limits, 
from $M_r=-20.4$ down to $M_r=-14.4$. This result is shown in 
\autoref{fig:HIHM_limits}, with the solid black line representing 
results for all galaxies, and colored lines for red/blue and 
central/satellite galaxies, respectively.
The total \hi\ mass based on the group catalog is plotted as solid circles with 
error bars in each panel, and the lower limit of \hi\ mass predicted for the Virgo 
EVCC sample is plotted as pink squares with upward arrows in the 
last panel. For comparison, the \hi\ mass from \cite{2020ApJ...894...92G}
is plotted as a horizontal dotted line in each panel. 

As expected, the \hi\ mass increases as the magnitude limit goes deeper, 
but the increase is significant only when the luminosity limit is brighter 
than $M_r\sim-18$. For limits deeper than $M_r\sim-18$, the 
total \hi\ mass estimated from the group catalog increases very little, 
while the mass estimated from the CLFs increases slowly, with a total increase 
of $\sim$0.1-0.2 dex. This result is true for all types of galaxies and 
for halos of different masses, except for satellites 
in low-mass halos where the \hi\ mass increases more rapidly 
with the limiting luminosity. This increase has little effect on 
the total \hi\ mass, as it is dominated by central galaxies. 
The total \hi\ mass from \cite{2020ApJ...894...92G} matches our results 
at a limiting luminosity that increases from $M_r\sim-17$ 
at \mhalo$\sim10^{12}$\msun\ to $M_r\sim-20$ at 
\mhalo$\sim3-5\times10^{13}$\msun, indicating that the ALFALFA 
datacubes may have missed an increasingly significant fraction of 
\hi\ emission from faint satellites which dominate the total \hi\ 
mass in massive halos. However, it is also possible that our 
estimator somehow overpredicts the {\hi} mass for satellite 
galaxies in massive clusters, if the gas-to-stellar mass ratio 
is lower for satellites in massive groups than for 
galaxies of similar optical properties in the general field
(see \autoref{sec:enviroment_bias} for more discussion). 
Clearly, deeper \hi\ surveys are needed to 
resolve the \hi\ gas content in satellite galaxies of massive halos. 
For Virgo Cluster, our estimated \hi\ mass (lower limits) 
increases with the limiting luminosity in the same way as the 
SDSS-based estimates, thus with a constant offset that should be 
attributed to the incompleteness of the EVCC galaxy catalog.

\subsection{The \hi-halo mass relation in theoretical studies}
\label{subsec:discussion_HI_simulation}

It is interesting to compare the \hi-to-halo mass relations obtained in 
this paper with theoretical predictions in the literature. 
In \autoref{fig:group_MHI_Mh} we plot the \hi-halo relation predicted 
by the L-GALAXIES semi-analytical model \citep[][]{2013MNRAS.434.1531F},
by the hydrodynamical simulation Illustris \citep[][]{2014MNRAS.444.1518V}, 
and by its successor IllustrisTNG \citep[][]{2018MNRAS.475..676S,
	2018MNRAS.480.5113M,2018MNRAS.477.1206N,2018MNRAS.475..624N,
	2018MNRAS.475..648P}.
Although all the models predict a positive correlation between \mhi\ and the 
halo mass, they over-predict the \hi\ mass to various degrees. 
In particular, at fixed halo mass the Illustris simulation predicts 
a total \hi\ mass that is a factor of 3-5 higher than the observed value, 
while the L-Galaxies model over-predicts the \hi\ mass for Milky Way-mass 
halos but a better match to the data at lower and higher halo masses. 
The TNG result falls in between the other two models. 
For central galaxies, the TNG over-predicts the \hi\ mass at all halo 
masses, while Illustris over-predicts the central \hi\ mass 
for low-mass halos and under-predicts it for high-mass halos. 
The L-Galaxies model predicts no \hi\ gas in centrals in halos 
with masses above $\sim10^{13}$\msun\ but too much \hi\ gas in centrals 
of Milky Way halos. These discrepancies must be related to the various 
physical recipes implemented in the simulations and the semi-analytic model 
to regulate gas-related processes. Our measurements of the \hi-to-halo mass 
relation, as well as the CHIMFs, for both the total galaxy population 
and for different sub-populations, can provide important constraints 
on theoretical models. 

\subsection{Possible biases in the \hi\ estimator due to environmental dependence}
\label{sec:enviroment_bias}

The fact that our model prediction for group galaxies matches observational
results (see \autoref{fig:group_MHI_Mh}) suggests that galaxies of similar
optical properties have similar {\hi} mass, irrespective of whether a galaxy
resides in dense regions or in the general field. This result supports 
the basic assumption behind our {\hi} estimator that there is a tight 
relation between the \hi\ gas content and other intrinsic properties 
of a galaxy. However, gas stripping effects can make this relation complex. 
On the one hand, the {\hi} gas may be removed without strongly affecting 
star formation on a timescale of $<1$ Gyr \citep{2017MNRAS.466.1275B}. 
In this case $NUV-r$ should be better than $u-r$ for estimating the 
\hi\ gas content, as the former is dominated by younger stellar populations
and so can more closely trace the change in star formation.
On the other hand, there are also evidences indicating that star formation 
can be enhanced in cluster galaxies that experience gas stripping 
\citep{2021A&A...652A.153R,2019MNRAS.487.4580R}. This, in addition to 
the fact that the xGASS sample is selected for the general population 
of galaxies and lacks very rich groups/clusters as pointed out by
\cite{2013MNRAS.436...34C}, may lead to bias in our {\hi} estimator,
particularly for galaxies that have recently experienced gas stripping. 
In other words, if galaxies in rich clusters like Virgo follow different 
scaling relations of \hi\ content from the average population 
(field galaxies), a bias can still be present in our \hi\ estimator
which is calibrated by galaxies mostly outside galaxy clusters.

In fact, \citet{2013MNRAS.429.2191Z} tested whether the \hi\ estimator 
of L12 exhibits any significant biases in dense environment by examining 
the difference between the estimated and observed $\log (\mhi/\mstar)$ 
for group/cluster galaxies as a function of the environment overdensity
$\ln(1+\delta)$ (see their Figure 3). They found the L12 estimator only 
slightly underestimates the \hi-to-stellar mass ratio by at most 0.1 dex 
on average, and this bias was caused by the \hi-rich galaxies rather than 
the \hi-poor galaxies in the group/clusters (thus opposite to the 
expectation for an environment-induced bias). Since our estimator
is motivated from and quite similar to the L12 estimator, we expect
our estimator to also provide unbiased \hi\ content for galaxies 
in different environments. Nevertheless, considering the different 
calibration samples adopted in L12 and in 
our work, we should still keep in mind the possible bias in our estimator,
especially when applying it to galaxies in dense regions. 
A complete and deep survey of \hi\ 21cm emission covering a full range 
of environment is needed if one were to accurately quantify the possible
dependence of an \hi\ estimator on environment.

\subsection{Limitations and Prospects}

Our \hi\ gas estimator is calibrated with the xGASS sample 
which is limited to nearby galaxies with $z<0.05$ and $\mstar>10^9 \msun$.
Therefore, in principle our estimator can be applied only to galaxy 
samples selected to these redshift and mass limits, although this 
estimator is shown to be able to reproduce the ALFALFA HIMF down to 
masses as low as \mhi$\sim10^7$\msun\ (see \autoref{fig:GAMA}).
For this reason all the results presented in this work for 
\mstar$<10^9$\msun\ should be taken in caution, and future studies 
with real \hi\ observations of galaxies at lower masses would 
be needed to verify our predictions, e.g. the conditional HIMFs 
at the low-mass end.

In the near future, next generation {H\sc{i}} surveys, such as WALLABY
\citep[the Widefield ASKAP L-band Legacy All-sky Blind surveY]{2020Ap&SS.365..118K} 
and LADUMA \citep[Looking at the distant universe with the MeerKAT Array]{2012IAUS..284..496H}
will provide unprecedented {H\sc{i}} data in the local and distant universe and 
improve our understanding of the {H\sc{i}} gas content of galaxies. However, the 
gap between {H\sc{i}} and optical surveys will remain for many years, and 
this is particularly true for low-mass and high-redshift galaxies. In the 
local universe, galaxies with $\log M_*/M_{\odot}>9.0$ 
already have relatively complete {\hi} surveys, such as xGASS and RESOLVE.
Although {\hi} surveys, such as ALFALFA, THINGS \citep[The H I Nearby 
Galaxy Survey]{2008AJ....136.2563W} and LITTLE THINGS \citep[Local Irregulars 
That Trace Luminosity Extremes, The HI Nearby Galaxy Survey]{2012AJ....144..134H},
have provided useful information about the {\hi} gas properties for 
lower mass galaxies, the low-mass galaxy samples obtained are still strongly 
biased for {\hi}-rich systems. For high-redshift galaxies, so far the only way 
to observe the \hi\ gas content is through damped Lyman $\alpha$ systems.
The estimator presented here, therefore, provides a way to predict and study 
\hi\ gas contents for large samples of optically selected galaxies 
that cover large ranges of galaxy properties and redshift.  
For instance, \citet{2021A&A...648A..25Z} have recently attempted to 
estimate the HIMF of galaxies at $z\sim1$ by applying a two-parameter 
\hi\ mass estimator calibrated with nearby galaxies to the DEEP2 galaxy sample. 
The results obtained in the paper demonstrate that such an approach is 
powerful and provides an important avenue to understand cold gas contents 
in galaxies and in galaxy systems.

\section{Summary} \label{sec:summary}

In this paper, we introduce a new {H\sc{i}} mass estimator in which the 
logarithm of the \hi-to-stellar mass ratio ($\log ($\mhi$/$\mstar$)$)
is modeled by a linear combination of four galaxy properties: 
surface stellar mass density ($\log\mu_\ast$),
color index ($u-r$), stellar mass ($\log M_\ast$) and concentration 
index ($\log (R_{90}/R_{50})$).
The variance of individual galaxies around the mean \hi\ mass is modeled 
by a Gaussian distribution. We calibrate the estimator with 
the xGASS sample, which is constructed to be representative for 
the {H\sc{i}} content of the local galaxy population. 
We constrain our model parameters using Bayesian inferences and 
including both detections and non-detections from the xGASS sample. 
Applying our estimator to mock catalogs that include the same selection 
effects as the ALFALFA survey, we are able to statistically reproduce 
the distributions of both the \hi\ mass and other galaxy properties of 
the real sample. We then apply our estimator to the SDSS spectroscopic 
galaxy sample to predict the \hi\ mass function (HIMF) of local galaxies, 
as well as the conditional \hi\ mass function (CHIMF) and the \hi-halo mass relation
(HIHM) for dark matter halos of different masses. 

Our main results can be summarized as follows.
\begin{itemize}
	\item The HIMF derived from the optical sample can be fitted with a single Schechter 
	function, but with a higher amplitude and a steeper slope at \hi\ masses below 
	a few $\times10^9$\msun\ when compared to the ALFALFA HIMF. 
	We show that this difference is essentially caused by the cosmic variance 
	of the local volume, which is corrected for our HIMF but not for the ALFALFA HIMF.
   \item We use our estimator to predict the CHIMFs for dark matter halos 
   with \mhalo$\ga10^{12}$\msun, for both the total galaxy population and sub-populations 
   of red/blue and central/satellite galaxies. A single Schechter function can well 
   describe all the CHIMFs at all halo masses, except for central galaxies where 
   a double Gaussian is preferred.  
   \item Although red galaxies dominate the galaxy population in massive halos, 
   red and blue galaxies have similar CHIMFs and thus contribute 
   similarly to the total \hi\ mass at given halo mass. This result 
   is true for all halo masses.
   \item Central galaxies contribute significantly to the total \hi\ mass only in low-mass 
   halos, while in massive halos the majority of \hi\ gas is associated with satellite galaxies. 
  \item The \hi\ mass of central galaxies shows very weak dependence on halo 
   mass, even when divided into red and blue sub-populations. This is quite different 
   from the total \hi\ mass of all galaxies which increases monotonically with increasing
   halo mass. 
   \item Current hydrodynamical simulations and semi-analytic models of galaxy 
   formation fail to reproduce the \hi-halo mass relation for both the central 
   population and the total population. 
\end{itemize}

\section*{Acknowledgments}

We are grateful to the anonymous referee for his/her comments 
which have helped improve our paper. XL thanks Kai Wang and Yangyao Chen for 
helpful discussions.
This work is supported by the National Key R\&D Program of China
(grant No. 2018YFA0404502), and the National Natural Science 
Foundation of China (grant Nos. 11821303, 11733002, 11973030, 
11673015, 11733004, 11761131004, 11761141012). 
This work has made use of the following softwares: Numpy \citep{2020Natur.585..357H}, Scipy \citep{2020SciPy-NMeth},
Matplotlib \citep{2007CSE.....9...90H}, emcee \citep{emcee}, Astropy \citep{2013A&A...558A..33A}, GetDist \citep{2019arXiv191013970L}, seaborn \citep{Waskom2021}, and h5py \citep{2021zndo...5585380C}.

Funding for the SDSS and SDSS-II has been provided by the Alfred P. Sloan Foundation, the Participating Institutions, the National Science Foundation, the U.S. Department of Energy, the National Aeronautics and Space Administration, the Japanese Monbukagakusho, the Max Planck Society, and the Higher Education Funding Council for England. The SDSS Web Site is http://www.sdss.org/.

The SDSS is managed by the Astrophysical Research Consortium for the Participating Institutions. The Participating Institutions are the American Museum of Natural History, Astrophysical Institute Potsdam, University of Basel, University of Cambridge, Case Western Reserve University, University of Chicago, Drexel University, Fermilab, the Institute for Advanced Study, the Japan Participation Group, Johns Hopkins University, the Joint Institute for Nuclear Astrophysics, the Kavli Institute for Particle Astrophysics and Cosmology, the Korean Scientist Group, the Chinese Academy of Sciences (LAMOST), Los Alamos National Laboratory, the Max-Planck-Institute for Astronomy (MPIA), the Max-Planck-Institute for Astrophysics (MPA), New Mexico State University, Ohio State University, University of Pittsburgh, University of Portsmouth, Princeton University, the United States Naval Observatory, and the University of Washington.

\bibliography{cite_list}{}
\bibliographystyle{aasjournal}

\end{document}